%% file: root_with_supplement.tex
\documentclass[lettersize,journal]{IEEEtran}
\input{preamble.tex}
\definecolor{cnavy}{RGB}{31,58,110}
\definecolor{cpurp}{RGB}{88,64,150}
\definecolor{cred}{RGB}{170,45,45}
\definecolor{cgreen}{RGB}{35,110,70}
\definecolor{cgray}{RGB}{95,95,105}
\definecolor{fnavy}{HTML}{1D3558}
\definecolor{fpanel}{HTML}{EDF3F9}
\definecolor{fbox}{HTML}{D6E0E8}
\definecolor{fchip}{HTML}{E3EEFA}
\definecolor{forange}{HTML}{CD915B}
\definecolor{fgray}{HTML}{5E6E82}
\definecolor{corange}{RGB}{204,137,74}

\usepackage[subtle]{savetrees}
\usepackage{titlesec}
\titlespacing*{\section}{0pt}{0.1\baselineskip}{0.1\baselineskip}
\titlespacing*{\subsection}{0pt}{0.1\baselineskip}{0.1\baselineskip}
\usepackage[dvipsnames]{xcolor}

\newcommand{\eqlabel}[1]{
  \text{ \setlength{\fboxsep}{0.1pt} \colorbox{Turquoise!0}{
      \makebox[2cm][r]{\strut\textit{#1}} }}}

\begin{document}
\title{Data Center Load Reductions: Attribution, Measurement Limits, and Estimation}
\author{Abdallah Alalem Albustami, Ahmad F.~Taha, Mohammad Alwahdanee, and Zuhaib Shakeel Muhammad\thanks{A.~Alalem Albustami and A.~F.~Taha are with the Department of Civil and Environmental Engineering, Vanderbilt University, Nashville, TN 37235, USA (e-mail: \{abdallah.b.alalem.albustami, ahmad.taha\}@vanderbilt.edu).}\thanks{M.~Alwahdanee is an independent researcher (e-mail: mohahmwah@gmail.com).}\thanks{Z.~S.~Muhammad is with Sibel Health, Seoul, South Korea (e-mail: zuhaib.shakeel@sibelhealth.com).}\vspace{-0.75cm}}
\maketitle

\begin{abstract}
Hundreds of gigawatts of data center loads are seeking interconnection in the United States given the massive growth of artificial intelligence and cloud computing. In recent years, several incidents of data centers reducing their loads by hundreds to thousands of megawatts have been documented, raising grid reliability concerns. U.S. regulators have since ordered reliability standards for the integration of computational loads, and the North American Electric Reliability Corporation (NERC) now calls on transmission owners to install disturbance recorders at these facilities to support root cause analysis. Such recorders usually observe only electrical quantities at the point of interconnection (POI). A sudden load reduction can be caused by a change in workload or a switch to backup supply.  This work shows when POI measurements can distinguish the two mechanisms. It is established that they can produce identical POI voltage and current in a quasi-static constant power model, and during the transient their separation depends on the reduction relative to measurement noise, the clearing time, the dynamics through which workload reaches the POI, and the recorder bandwidth. When POI data are insufficient, an estimator that uses workload data is proposed to determine how much of the facility stayed connected. The results of this work identify the measurements and grid information utilities need to determine why data center load dropped.
\end{abstract}


\section{Introduction}\label{sec:intro}
\lettrine[lraise=0.1,nindent=0em,slope=-.5em]{W}{ith} the rapid growth of artificial intelligence, cloud computing, and the emergence of large language models, hundreds of gigawatts of data center load are seeking interconnection in the United States, including about 410~GW of large loads in Texas alone as of March 2026, roughly 87\% of them data centers~\cite{ercot_senate_2026}, making them increasingly critical for grid reliability. In July 2024, a transmission fault in the Eastern Interconnection was followed within seconds by a reduction of approximately 1,500~MW in data center load~\cite{nerc_ir_2025}. The North American Electric Reliability Corporation (NERC) has since documented additional events in which a gigawatt or more of computational load dropped off the power system within seconds, issued a Level~3 essential action alert on the modeling, study, commissioning, protection, and instrumentation of such loads~\cite{nerc_l2_2025,nerc_l3_2026}, and published a reliability guideline for emerging large loads~\cite{nerc_rg_2026}. In July 2026 the Federal Energy Regulatory Commission directed NERC to file reliability standards for the integration of computational loads by December~31, 2026~\cite{ferc_rd26_2026}.

These reductions have several causes. The facility's own protection may disconnect equipment or switch it to backup supply (such as uninterruptible power supply (UPS) batteries or onsite generators) or the computing workload itself can drop, for example when a training job pauses or servers throttle~\cite{nerc_ir_2025,nerc_l2_2025}.  This work considers these two mechanisms. Current limiting, reactive current priority, and auxiliary load changes also change the POI current; the first two are part of the all-connected response when the facility model represents them, and a trip of auxiliary load appears as a disconnection.

Transmission operators and planners need to understand data center responses to assess ride-through performance, validate load models, and analyze disturbances~\cite{nerc_rg_2026}. As these facilities grow in size and number, rapid changes in workload can have greater consequences for grid operation~\cite{nerc_wp_2025}. NERC also mentions shared computing workloads as a mechanism through which a disturbance at one facility can trigger demand reductions at others~\cite{nerc_rg_2026}. To support event analysis, NERC calls on transmission owners to install disturbance recorders at computational facilities~\cite{nerc_l3_2026}. Interpreting these records requires understanding which changes in the measured response can arise from workload variation and which can be explained by transfer to backup supply or protection action, which is the problem addressed herein.

Facility event logs can show that a UPS transferred to backup supply, but not how much grid current was lost when the workload also changed, and computational loads are not yet required to share them. Attribution from synchronized POI and workload data provides an independent and quantitative check.

Specifically, the objective of this work is threefold. \textit{(i)} Determine when POI voltage and current can distinguish these mechanisms and how this depends on reduction magnitude, measurement noise, protection clearing time, workload dynamics, and recorder bandwidth. \textit{(ii)} Identify the additional workload and grid information needed when POI measurements are insufficient. \textit{(iii)} Develop an estimator that uses these data to reconstruct the current expected with all equipment connected and estimate how much of that response remains at the POI.

\vspace{0.1cm}
\noindent\textbf{Related Work.}
Measurement-based load modeling estimates from bus records parameters such as the fraction of electronic load that disconnects during a disturbance~\cite{wecc_cmpldw,nerc_loadmodel_2016,kosterev2008}, although several parameter combinations can reproduce the same record~\cite{ma2008}, and fault-induced delayed voltage recovery analyses explain post-fault load loss through motor stalling and tripping~\cite{nerc_fidvr_2009}. Both attribute load changes to equipment rather than to demand, which for computational loads can vary on comparable time scales~\cite{choukse2025} and has been linked to wide-area oscillations~\cite{ko2026,mishra2025}. Studies of data center dynamics and ride-through treat demand changes and protective disconnection as distinct~\cite{sun2022,kwon2025,chen2025,xie2025enhancing,liang2026,colangelo2026}. Demand response baselines and load disaggregation infer counterfactual demand or separate aggregate consumption~\cite{coughlin2009,hart1992} at much slower time scales. Whether POI measurements can distinguish workload changes from backup transfer or disconnection, and what else is needed when they cannot, is what we call the attribution problem.

\vspace{0.1cm}
\noindent\textbf{Contributions.}
The main contributions are as follows.
\begin{itemize}[leftmargin=*]

\item Protection disconnection and workload reduction are shown to produce identical POI voltage and current in a quasi-static constant power model. A dynamic separation condition then establishes when their transient responses can be distinguished and shows how separation depends on reduction magnitude, measurement noise, clearing time, workload dynamics, and recorder bandwidth.

\item A response model fitted directly to measured POI voltage can reproduce the grid voltage-current relation instead of the facility response and therefore mask a disconnection. Reconstructing the source voltage, which requires the grid impedance, removes this relation by construction and gave smaller errors in the tests; inputs and regularization selected on calibration records also removed the failure with POI voltage.

\item When POI measurements are insufficient, workload data are used to predict the all-connected response, and an estimator determines how much of it remains in the measured POI current and how much load was lost to disconnection and to workload change, withholds results when the prediction is not valid, and flags a current that is not a scaled copy of the prediction.

\item The results give practical requirements for attribution. Workload data must represent computational demand independently of the grid connection state, the recorder must retain enough of the transient response to distinguish the mechanisms, and zone-level workload data remove the allocation ambiguity that remains with aggregate workload data.
\end{itemize}

\vspace{0.1cm}
\noindent\textbf{Organization.} Section~\ref{sec:model} develops the facility model and the attribution problem. Section~\ref{sec:identifiability} states what terminal measurements identify. Section~\ref{sec:estimator} develops the estimator using workload data and its extensions. Section~\ref{sec:requirements} collects the data requirements, Section~\ref{sec:case_studies} tests the estimator on PSCAD models of three facilities with complete and partial disconnections, and Section~\ref{sec:conclusion} concludes the paper. Derivations used in the paper are in the supplementary material (SM).

\begin{figure*}[t]
    \centering
    \includegraphics[width=0.9\linewidth]{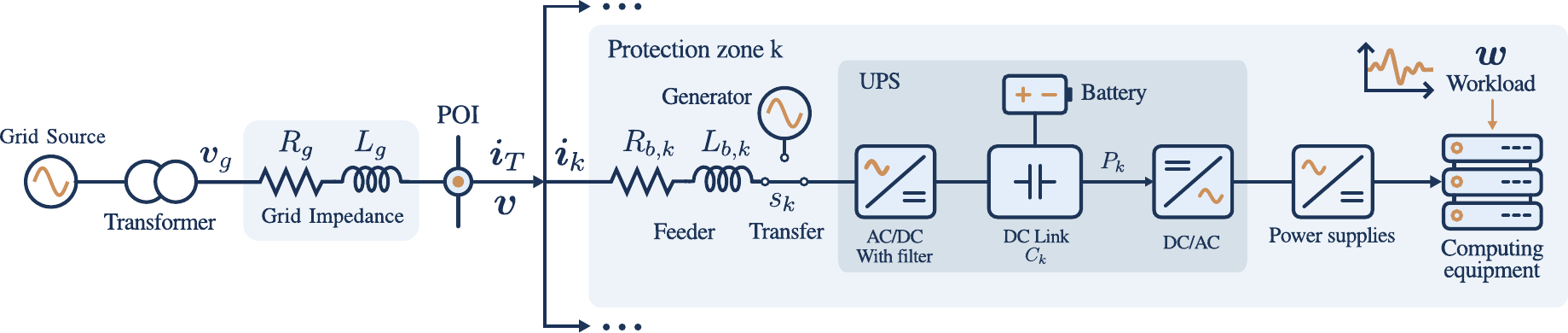}
    \caption{Grid connection and power supply components of a computational facility, with one protection zone shown in detail.}
    \label{fig:facility_architecture}
    \vspace{-0.45cm}
\end{figure*}

\section{Facility Model and the Attribution Problem}\label{sec:model}

A computational load (such as a data center or a cryptocurrency mining facility) connects to the grid through transformers, circuit breakers, and feeders that deliver power to the computing equipment. Cooling and other auxiliary equipment also contribute to the facility's demand. Fig.~\ref{fig:facility_architecture} illustrates this power supply arrangement and the simplified model used here, which groups the computational equipment into zones, each represented by a feeder, a converter, a DC link, and the computing demand it supplies.

\subsection{Facility and grid model}\label{subsec:facility_model}

The grid is represented by a Th\'{e}venin source voltage $\m v_g$ behind a resistance $R_g$ and inductance $L_g$. The facility draws current $\m i_T$ at the POI voltage $\m v$. Voltages and currents are expressed in a power invariant synchronous $dq$ frame rotating at nominal angular frequency $\omega_0$, so active power is $\m v^\top\m i_T$. The voltage balance between the source and POI is
\begin{equation}\label{eq:grid}
\m v=\m v_g-R_g\m i_T-L_g\frac{d\m i_T}{dt}-\omega_0L_g\m J\m i_T,
\end{equation}
where $\m J=\left[\begin{smallmatrix}0&-1\\1&0\end{smallmatrix}\right]$. The source voltage $\m v_g$ represents external grid conditions, including voltage sags and phase jumps, and is treated as an input to the model. The POI voltage $\m v$, however, changes with the facility current through~\eqref{eq:grid}.

The facility consists of $N$ protection zones, each a set of equipment that disconnects together (such as the load behind one UPS or one feeder breaker). Zone $k$ is connected through a series feeder $(R_{b,k},L_{b,k})$ and draws its current through a converter with filter $(R_{f,k},L_{f,k})$, a DC link of capacitance $C_k$, and a controller with an outer DC voltage loop and an inner current loop (see Fig.~\ref{fig:facility_architecture}). With $\m v_k$ the zone terminal voltage, $\m i_k$ the zone current, $v_{dc,k}$ the DC link voltage, $\m u_k$ the converter voltage, $P_k(t)$ the power drawn from the DC link by the computational equipment, and $\m z_k$ the controller states, the dynamics of zone $k$ are then given by

\vspace{-0.2cm}
\begin{subequations}\label{eq:zone}\small
\begin{alignat}{2} 
&\hspace{-0.6cm}\eqlabel{\textbf{Feeder}} \;\;\;\;
&&\m v_k=\m v-R_{b,k}\m i_k-L_{b,k}\dot{\m i}_k-\omega_0L_{b,k}\m J\m i_k,
\label{eq:feeder}\\
&\hspace{-0.6cm}\eqlabel{\textbf{Filter}}
&&L_{f,k}\dot{\m i}_k=\m v_k-R_{f,k}\m i_k-\omega_0L_{f,k}\m J\m i_k-\m u_k,
\label{eq:zone_current}\\
&\hspace{-0.6cm}\eqlabel{\textbf{DC link}}
&&C_kv_{dc,k}\dot v_{dc,k}=\m u_k^\top\m i_k-P_k(t),
\label{eq:zone_dclink}\\
&\hspace{-0.6cm}\eqlabel{\textbf{Controller}}
&&\dot{\m z}_k=\m\phi_k(\m x_k,\m v_k,\m i_k,v_{dc,k}),
\label{eq:zone_controller_state}\\
&\hspace{-0.6cm}\eqlabel{\textbf{Control output}}
&&\m u_k=\m\kappa_k(\m x_k,\m v_k,\m i_k,v_{dc,k}),
\label{eq:zone_control}
\end{alignat}
\end{subequations}
where $\m\phi_k$ and $\m\kappa_k$ are the controller state and output maps. Workload enters through DC link balance~\eqref{eq:zone_dclink}. A change in $P_k$ first moves $v_{dc,k}$, the outer loop then moves the current reference, and the inner loop moves $\m i_k$. DC link and converter control dynamics can hence determine how quickly workload change reaches terminal current.

For each protection zone, $s_k\in\{0,1\}$ denotes its connection state and $m_k=\mathcal F_k[\|\m v_k\|]$ denotes filtered terminal voltage magnitude used by protection, where $\mathcal F_k$ is a causal measurement filter (such as an RMS or low-pass filter). Zone $k$ disconnects when $m_k$ remains below the threshold $\lambda_k$ for at least the delay $\tau_k$,
\begin{equation}\label{eq:protection}
m_k(t')<\lambda_k\ \text{for all}\ t'\in[t-\tau_k,t]\ \Longrightarrow\ s_k^+=0,
\end{equation}
after which $\m i_k=0$ while stored energy or backup supply supports the downstream computational equipment. Input filter capacitors keep drawing reactive current after a disconnection, the model omits them and Section~\ref{subsec:static} reports their effect. Let $\m x$ collect the continuous electrical and controller states of all zones as well as any states of protection filters $\mathcal F_k$. The connected set is $\mathcal S=\{k\mid s_k=1\}$, POI current is $\m i_T=\sum_{k\in\mathcal S}\m i_k$, and workload is $\m w=\col(P_1,\ldots,P_N)$ with aggregate defined as $w_T=\sum_{k=1}^N P_k$. Combining the grid and zone equations gives the hybrid facility model below, whose dynamics depend on the connected set $\mathcal S$.
\begin{equation}\label{eq:hybrid}
\dot{\m x}=\m f_{\mathcal S}(\m x,\m v_g,\m w),\qquad \m i_T=\m h_{\mathcal S}(\m x),
\end{equation}
with $\mathcal S$ changing at protection events. Disconnection is modeled as an instantaneous drop in the zone's grid current to zero. All other continuous states retain their values at that instant.

\subsection{All-connected response and participation}\label{subsec:participation}

To distinguish the load reduction mechanism, we must know how much of the facility's response is still present at the POI. This requires a reference to compare with. For a given source voltage $\m v_g(\cdot)$, workload $\m w(\cdot)$, and pre-disturbance continuous state with all zones connected, the \textit{all-connected response} $\m i^{(N)}(\cdot)$ is simply the POI current of~\eqref{eq:hybrid} with $\mathcal S(t)\equiv\{1,\ldots,N\}$ (i.e., all zones connected). For a specified disconnection time $t_d$, the partial state response $\m i^{(\mathcal S)}(\cdot)$ follows the same all-connected trajectory up to $t_d$, applies the protection to the zones outside $\mathcal S$, and then holds the connected set at $\mathcal S$.

The all-connected response is referenced to the source voltage rather than the POI voltage since POI voltage itself changes when equipment disconnects through~\eqref{eq:grid}. Over some window $\mathcal W$ of $n$ samples, with $\langle\m a,\m b\rangle_{\mathcal W}=\sum_{t\in\mathcal W}\m a_t^\top\m b_t$ and $\langle\m i^{(N)},\m i^{(N)}\rangle_{\mathcal W}>0$, define the electrical participation as the constrained least-squares coefficient
\begin{equation}\label{eq:participation}
a_e(\mathcal W)=\arg\min_{a\in[0,1]}\sum_{t\in\mathcal W}\|\m i_{T,t}-a\m i^{(N)}_t\|_2^2.
\end{equation}
Here $a_e$ is the unconstrained least-squares ratio limited to $[0,1]$. It is the fraction of the all-connected current response present in the measured POI current in this constrained least-squares sense and is generally not equal to the connected nominal power fraction because the POI voltage changes when equipment disconnects. 

For a connected set $\mathcal S$, define the connected nominal power fraction as
\begin{equation}\label{eq:capacity_count}
c(\mathcal S)=\frac{\sum_{k\in\mathcal S}P_k^0}{\sum_{k=1}^NP_k^0},
\end{equation}
where $P_k^0>0$ is the nominal power of zone $k$.
In quasi-static constant power operation at unity power factor, with zone powers proportional to $P_k^0$, negligible feeder drops, negligible phase difference between the all-connected and partially connected POI voltages, POI voltage magnitudes $\bar v^{(N)}$ and $\bar v^{(\mathcal S)}$ that are constant over the window, and operation on the high-voltage branch, the two quantities are related by
\begin{equation}\label{eq:ae_capacity}
a_e(\mathcal W)\approx c(\mathcal S)\,\frac{\bar v^{(N)}}{\bar v^{(\mathcal S)}}\le c(\mathcal S).
\end{equation}
The inequality holds because reducing the connected power raises the POI voltage on the high-voltage branch (SM, Section~S1). We note here that the above assumptions are used only to obtain the closed form in~\eqref{eq:ae_capacity} and are not required by the implementation which compares the measured POI current with the dynamic all-connected response. In a general facility, feeder drops, reactive response, unequal loading, and converter dynamics may alter the relationship between electrical participation and connected nominal power.

\subsection{Information sets and the attribution problem}\label{subsec:information}

The POI measurements consist of terminal voltage and current sampled at the recorder rate with measurement noise. When these measurements do not help distinguish the load reduction mechanism (protection disconnection or workload change), we need facility data of the computing load for attribution. For a data center, this can be an estimate of the power demanded by the servers or accelerators that remains available when equipment disconnects from the grid or transfers to backup supply. The workload data must represent the computational power demand $P_k(t)$ or an estimate of it independently of the zone's grid connection state over the attribution window. Data that changes directly because the zone disconnects from the grid are uninformative since such a signal cannot make the distinction. This requirement is referred to below as \textit{workload data validity}. Based on the available measurements and workload data, first define the information sets
\begin{subequations}\label{eq:information_sets}
\begin{align}
\mathcal I_0&=\{\m v,\m i_T\},\qquad
\mathcal I_1=\mathcal I_0\cup\{w_T^{\rm obs}\},\label{eq:I01}\\
\mathcal I_2&=\mathcal I_1\cup\{\mathcal C\},\qquad
\mathcal I_3=\mathcal I_0\cup\{P_k^{\rm obs}\}_{k=1}^N\cup\{\mathcal C\},\label{eq:I23}
\end{align}
\end{subequations}
where $P_k^{\rm obs}$ is the observed workload of zone $k$, $w_T^{\rm obs}=\sum_kP_k^{\rm obs}$ is the aggregate workload data, and $\mathcal C$ contains the equipment class and nominal power of each zone. Thus, $\mathcal I_0$ contains only the POI measurements, $\mathcal I_1$ adds aggregate workload data, $\mathcal I_2$ also includes the equipment class and nominal power of each zone, and $\mathcal I_3$ replaces aggregate workload with zone-level workload data. The grid impedance is treated separately because it is network information not facility data.

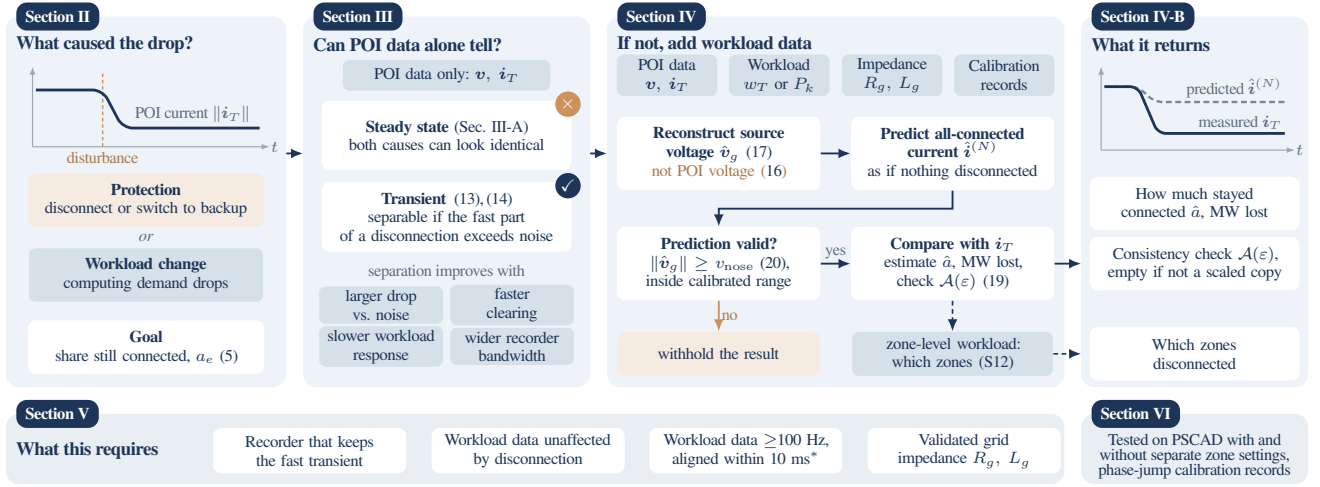
\begin{figure*}[t]
\centering
\resizebox{0.95\textwidth}{!}{\hypersetup{linkcolor=fnavy}%
\begin{tikzpicture}[
  x=1cm, y=1cm, font=\scriptsize, text=fnavy,
  tab/.style={rectangle, rounded corners=1.6mm, fill=fnavy, text=white,
    font=\scriptsize\bfseries, minimum height=4.2mm, inner xsep=4pt, inner ysep=0pt, anchor=west},
  head/.style={font=\footnotesize\bfseries, text=fnavy, anchor=west, inner sep=0pt},
  card/.style={fill=white, draw=none, rounded corners=1.2mm, align=center, inner sep=2pt},
  chip/.style={fill=fbox, draw=none, rounded corners=1mm, align=center, inner sep=1.5pt},
  ochip/.style={fill=forange!18, draw=none, rounded corners=1.2mm, align=center, inner sep=2pt},
  flow/.style={-{Latex[length=1.8mm,width=1.5mm]}, fnavy, line width=0.8pt},
  oflow/.style={-{Latex[length=1.8mm,width=1.5mm]}, forange, line width=0.8pt},
  dflow/.style={-{Latex[length=1.8mm,width=1.5mm]}, fnavy, line width=0.7pt, dash pattern=on 2pt off 1.5pt},
  okmark/.style={circle, inner sep=0pt, minimum size=3.8mm, font=\scriptsize\bfseries, text=white},
  lbl/.style={font=\scriptsize, text=fgray, inner sep=1pt},
  axis/.style={fgray!60, line width=0.5pt, -{Latex[length=1.2mm]}},
  p1box/.style={text width=34mm, minimum height=8mm},
  p2box/.style={card, text width=36mm, minimum height=10.4mm},
  p2chip/.style={chip, text width=17.5mm, minimum height=5mm},
  p3chip/.style={chip, text width=14mm, minimum height=6.2mm},
  p3box/.style={card, text width=29mm, minimum height=11mm},
  p3out/.style={text width=29mm, minimum height=6.6mm},
  p4box/.style={card, text width=30mm, minimum height=8mm},
  reqchip/.style={card, text width=27.5mm, minimum height=6.8mm},
]
\fill[fpanel, rounded corners=2mm] (0.00,1.45) rectangle (4.20,7.00);
\fill[fpanel, rounded corners=2mm] (4.45,1.45) rectangle (8.75,7.00);
\fill[fpanel, rounded corners=2mm] (9.00,1.45) rectangle (15.85,7.00);
\fill[fpanel, rounded corners=2mm] (16.10,1.45) rectangle (19.50,7.00);
\fill[fbox!55, rounded corners=2mm] (0.00,0.00) rectangle (15.85,1.05);
\fill[fbox!55, rounded corners=2mm] (16.10,0.00) rectangle (19.50,1.05);

\node[tab] at (0.15,7.00)  {Section~\ref*{sec:model}};
\node[tab] at (4.60,7.00)  {Section~\ref*{sec:identifiability}};
\node[tab] at (9.15,7.00)  {Section~\ref*{sec:estimator}};
\node[tab] at (16.25,7.00) {Section~\ref*{subsec:certificate}};
\node[tab] at (0.15,1.05)  {Section~\ref*{sec:requirements}};
\node[tab] at (16.25,1.05) {Section~\ref*{sec:case_studies}};
\node[head] at (0.15,6.58)  {What caused the drop?};
\node[head] at (4.60,6.58)  {Can POI data alone tell?};
\node[head] at (9.15,6.58)  {If not, add workload data};
\node[head] at (16.25,6.58) {What it returns};

\draw[axis] (0.40,5.05) -- (3.95,5.05) node[right, lbl, inner sep=0.5pt] {$t$};
\draw[axis] (0.40,5.05) -- (0.40,6.25);
\draw[forange, line width=0.6pt, dash pattern=on 1.5pt off 1.2pt] (1.45,5.05) -- (1.45,6.15);
\node[lbl, text=forange!85!black, anchor=north] at (1.45,5.02) {disturbance};
\draw[fnavy, line width=1.2pt, rounded corners=1.5mm] (0.45,5.90) -- (1.45,5.90) -- (1.75,5.33) -- (3.80,5.33);
\node[lbl, anchor=south] at (2.80,5.37) {POI current $\|\m i_T\|$};
\node[ochip, p1box] at (2.10,4.25) {\textbf{Protection}\\disconnect or switch to backup};
\node[lbl, font=\scriptsize\itshape] at (2.10,3.69) {or};
\node[chip,  p1box] at (2.10,3.13) {\textbf{Workload change}\\computing demand drops};
\node[card,  p1box] at (2.10,2.05) {\textbf{Goal}\\share still connected, $a_e$~\eqref{eq:participation}};

\node[chip, text width=30mm, minimum height=4.6mm] at (6.60,6.12) {POI data only: $\m v,\ \m i_T$};
\node[p2box] (ss) at (6.60,5.22) {\textbf{Steady state}~(Sec.~\ref{subsec:static})\\both causes can look identical};
\node[okmark, fill=forange] at ([xshift=-0.6mm,yshift=-0.6mm]ss.north east) {$\times$};
\node[p2box] (tr) at (6.60,4.00) {\textbf{Transient}~\eqref{eq:separation},\,\eqref{eq:d_lower}\\separable if the fast part\\of a disconnection exceeds noise};
\node[okmark, fill=fnavy] at ([xshift=-0.6mm,yshift=-0.6mm]tr.north east) {$\checkmark$};
\node[lbl] at (6.60,3.17) {separation improves with};
\node[p2chip] at (5.62,2.68) {larger drop\\vs.\ noise};
\node[p2chip] at (7.58,2.68) {faster\\clearing};
\node[p2chip] at (5.62,2.06) {slower workload\\response};
\node[p2chip] at (7.58,2.06) {wider recorder\\bandwidth};

\node[p3chip] at (9.90,6.12)  {POI data\\$\m v,\ \m i_T$};
\node[p3chip] at (11.58,6.12) {Workload\\$w_T$ or $P_k$};
\node[p3chip] at (13.27,6.12) {Impedance\\$R_g,\ L_g$};
\node[p3chip] at (14.95,6.12) {Calibration\\records};

\node[p3box] (R) at (10.675,4.95) {\textbf{Reconstruct source}\\\textbf{voltage} $\hat{\m v}_g$~\eqref{eq:source_reconstruction}\\{\color{forange!85!black}not POI voltage~\eqref{eq:q_inversion}}};
\node[p3box] (M) at (14.175,4.95) {\textbf{Predict all-connected}\\\textbf{current} $\hat{\m i}^{(N)}$\\as if nothing disconnected};
\node[p3box] (F) at (10.675,3.30) {\textbf{Prediction valid?}\\$\|\hat{\m v}_g\|\ge v_{\rm nose}$~\eqref{eq:nose},\\inside calibrated range};
\node[p3box] (C) at (14.175,3.30) {\textbf{Compare with} $\m i_T$\\estimate $\hat a$, MW lost,\\check $\mathcal A(\varepsilon)$~\eqref{eq:interval}};
\node[ochip, p3out] (W) at (10.675,1.95) {withhold the result};
\node[chip,  p3out] (Z) at (14.175,1.95) {zone-level workload:\\which zones~\eqref{eq:subset_family}};

\draw[flow] (R.east) -- (M.west);
\draw[flow] (M.south) -- ++(0,-0.27) -| (F.north);
\draw[flow] (F.east) -- node[lbl, above, yshift=0.2mm] {yes} (C.west);
\draw[oflow] (F.south) -- node[lbl, right, text=forange!85!black] {no} (W.north);
\draw[dflow] (C.south) -- (Z.north);

\draw[axis] (16.40,5.00) -- (19.30,5.00) node[right, lbl, inner sep=0.5pt] {$t$};
\draw[axis] (16.40,5.00) -- (16.40,6.25);
\draw[fnavy!60, line width=1.0pt, dash pattern=on 2.5pt off 1.5pt, rounded corners=1.2mm]
  (16.45,5.95) -- (16.95,5.95) -- (17.20,5.72) -- (19.15,5.72);
\draw[fnavy, line width=1.2pt, rounded corners=1.2mm]
  (16.45,5.95) -- (16.95,5.95) -- (17.25,5.25) -- (19.15,5.25);
\node[lbl, anchor=south east] at (19.15,5.74) {predicted $\hat{\m i}^{(N)}$};
\node[lbl, anchor=south east] at (19.15,5.27) {measured $\m i_T$};
\node[p4box] at (17.80,4.20) {How much stayed\\connected $\hat a$, MW lost};
\node[p4box] at (17.80,3.30) {Consistency check $\mathcal A(\varepsilon)$,\\empty if not a scaled copy};
\node[p4box] at (17.80,1.95) {Which zones\\disconnected};

\draw[flow] (4.20,4.95) -- (4.45,4.95);
\draw[flow] (8.75,4.95) -- (9.00,4.95);
\draw[flow] (C.east) -- (16.10,3.30);
\draw[dflow] (Z.east) -- (16.10,1.95);

\node[head] at (0.15,0.50) {What this requires};
\node[reqchip] at (4.55,0.50)  {Recorder that keeps\\the fast transient};
\node[reqchip] at (7.82,0.50)  {Workload data unaffected\\by disconnection};
\node[reqchip] at (11.08,0.50) {Workload data $\ge$100~Hz,\\aligned within 10~ms$^{*}$};
\node[reqchip] at (14.35,0.50) {Validated grid\\impedance $R_g,\ L_g$};
\node[anchor=center, align=center, text=fnavy] at (17.80,0.42)
  {Tested on PSCAD with and\\[-1.5pt]without separate zone settings,\\[-1.5pt]phase-jump calibration records};
\end{tikzpicture}}
\caption{Overview of the paper. Section~\ref{sec:identifiability} shows what POI data alone can distinguish, Section~\ref{sec:estimator} adds workload data to estimate participation, Section~\ref{sec:requirements} lists data requirements, and Section~\ref{sec:case_studies} tests the approach on benchmark systems. $^{*}$From the tested cases.}
\label{fig:overview}\vspace{-0.4cm}
\end{figure*}

The attribution problem is then the following. Given an information set $\mathcal I_j$, any available grid impedance information, a window $\mathcal W$ within a disturbance, and a residual tolerance $\varepsilon$, estimate the participation, or the connected set, and the load lost to each mechanism. If the data are outside the calibrated range, report that condition instead of an attribution. If the measured current is not a scaled copy of the all-connected response within $\varepsilon$, report that condition with the estimate, because the single-participation model then does not describe the facility. Fig.~\ref{fig:overview} shows an overview of this work and a summary of the results presented in each section.

\section{What POI Measurements Can Distinguish}\label{sec:identifiability}

This section first shows that $\mathcal I_0$ cannot identify the mechanism in the quasi-static case. It then characterizes dynamic separation for a fixed grid disturbance when the transient terminal responses of the two mechanisms differ.

\subsection{Identical POI response in the constant power case}\label{subsec:static}
In quasi-static constant power operation (i.e., neglecting electrical transients and treating power demand as independent of voltage), a reduction in POI current can result either from less equipment remaining connected or from lower computational demand, and terminal measurements alone may not reveal which occurred. To show this, suppose the facility current can be written as
\begin{equation}\label{eq:separable}
\m i_T=\eta\,\mathcal G(\m v,u),
\end{equation}
where $\eta\in[0,1]$ is a multiplicative electrical connection factor, $u$ is a scalar demand, and $\mathcal G$ is a causal map from the terminal voltage history and the demand to the facility current. If $\mathcal G$ is positively homogeneous of degree one in demand, so that $\mathcal G(\m v,cu)=c\,\mathcal G(\m v,u)$, then the pairs $(\eta,u)$ and $(c\eta,u/c)$ give the same facility current for any terminal voltage history and any $c>0$ such that $c\eta\in[0,1]$ and $u/c$ is admissible. Under the same source voltage, the grid equation~\eqref{eq:grid} then gives the same POI voltage, so the two pairs produce identical POI voltage and current. For example, a record consistent with $\eta=0.8$ at demand $u$ is equally consistent with $\eta=1$ at demand $0.8u$, provided both operating points are admissible. No method using only $\mathcal I_0$ can distinguish these cases from the terminal data alone. The positive homogeneity assumption is used here simply to obtain the exact result above and is not required by the implementation in case studies.

The result also assumes that a disconnection removes a fraction of the whole current vector. In the PNNL models considered in Section~\ref{sec:case_studies}, a complete cessation (in which all converters stop drawing grid current) instead leaves a current that is more than 99\% reactive (0.07--0.30~$\pu$), consistent with input filters that stay energized, whereas after a workload reduction the remaining current is 28--45\% reactive. These models therefore show a reactive signature of complete cessation but not of partial disconnection (40\% after a cessation on part of the phases, 14\% after a support-branch trip), and the ratio of the reactive to the active power change alone does not separate the mechanisms.

\subsection{Dynamic separation}\label{subsec:dynamic_sep}

During a disturbance, the facility is not quasi-static because disconnection can remove grid current within the clearing time of the protection whereas a workload change reaches the POI through the DC link and converter controls. Their transient responses can therefore differ even when their quasi-static responses are identical.
To see this, the two mechanisms should be compared under the same source voltage trajectory. First, fix a window $\mathcal W$ of $n$ samples beginning at the disturbance, and let $\m y\in\RR^{4n}$ include the sampled POI voltage and current deviations from their pre-disturbance trajectories after whitening (scaling and decorrelating) by the corresponding sensor noise covariance so that $\m e\sim\mathcal N(\m 0,\m I)$. Let $\mathcal U$ be a closed set of admissible workload deviations. A general first-order representation on the window is
\begin{subequations}\label{eq:hypotheses}
\begin{align}
\mathcal H_0&\quad \m y=\m b_N+\mathcal P_N\m p+\m e,\qquad \m p\in\mathcal U,\label{eq:H0}\\
\mathcal H_1&\quad \m y=\m b_{\mathcal S}+\mathcal P_{\mathcal S}\m p'+\m e,\qquad \m p'\in\mathcal U.\label{eq:H1}
\end{align}
\end{subequations}
Here $\mathcal H_0$ represents the all-connected case and $\mathcal H_1$ the transition to $\mathcal S$. The vectors $\m b_N$ and $\m b_{\mathcal S}$ are the corresponding terminal responses to the fixed grid disturbance with the workload held at its pre-disturbance value. The operators $\mathcal P_N$ and $\mathcal P_{\mathcal S}$ map admissible workload deviations to the terminal measurements under the two cases. The noiseless response sets are
\begin{equation}\label{eq:response_sets}
\mathcal M_N=\m b_N+\mathcal P_N\mathcal U,\qquad
\mathcal M_{\mathcal S}=\m b_{\mathcal S}+\mathcal P_{\mathcal S}\mathcal U.
\end{equation}
The two mechanisms are distinguishable without noise on $\mathcal W$ if these sets do not intersect. Their whitened separation distance is 
\begin{equation}\label{eq:general_separation}
d_{\rm gen}(\mathcal S)=\inf_{\m p,\m p'\in\mathcal U}\|\m b_{\mathcal S}+\mathcal P_{\mathcal S}\m p'-\m b_N-\mathcal P_N\m p\|_2.
\end{equation}
A closed form of this distance can be obtained if every variation in POI voltage and current caused by workload changes after disconnection can also be produced by workload changes with all zones connected.

Now let $\m s_{\mathcal S}=\m b_{\mathcal S}-\m b_N$ denote the effect of the disconnection with the workload held fixed. Take $\mathcal U$ to be a linear subspace and assume $\mathcal P_{\mathcal S}\mathcal U\subseteq\mathcal P_N\mathcal U$, which holds in particular when $\mathcal P_N=\mathcal P_{\mathcal S}$. With $\m\Pi$ the orthogonal projector onto $\mathcal P_N\mathcal U$, applying $\m I-\m\Pi$ removes the components that workload changes can produce, and~\eqref{eq:general_separation} reduces to
\begin{equation}\label{eq:separation}
d(\mathcal S)=\|(\m I-\m\Pi)\m s_{\mathcal S}\|_2,
\end{equation}
the part of the disconnection effect that no workload change can reproduce (SM, Section~S2). Without noise, the mechanisms are distinguishable when $d(\mathcal S)>0$, and if $d(\mathcal S)=0$, workload variation can reproduce the entire disconnection effect. With noise, and assuming the two explanations are equally likely, a likelihood ratio test on the projected measurement selects whichever predicted response is closer and has error probability $P_e=Q(d(\mathcal S)/2)$, where $Q(z)$ is the probability that a standard normal variable exceeds $z$. Because $d(\mathcal S)$ is measured in units of noise, $P_e\le\alpha$ if and only if $d(\mathcal S)\ge2Q^{-1}(\alpha)$ for $0<\alpha<1/2$.

Considering only the current measurements gives us a simple lower bound on $d(\mathcal S)$ that shows what controls it. A workload change reaches the POI current only through the DC link and converter controls, so it can only produce changes up to some frequency $\omega_c$, set by the slower of workload and bandwidth $\omega_p$ of that path. A disconnection removes current within the clearing time so it also produces faster changes. Let $\Delta\m i_{\rm fast}$ denote the part of the disconnection's effect on the POI current at frequencies between $\omega_c$ and the recorder bandwidth $\omega_r$, that is, the part no workload change can produce but the recorder still captures. If the current measurement noise is independent from sample to sample, with standard deviation $\sigma_i$, and independent of the voltage measurement noise, then we get
\begin{equation}\label{eq:d_lower}
d(\mathcal S)\ge\frac{\|\Delta\m i_{\rm fast}\|_2}{\sigma_i}.
\end{equation}
This means the separation is at least the size of the fast part of the disconnection measured in units of current noise.

The clearing time is what controls how large that fast part is. To see this, suppose protection removes a current $\Delta$ over a time $\tau$. Up to a frequency of about $2/\tau$, the change looks like an instantaneous step, with spectrum magnitude $\Delta/\omega$, and above that frequency it falls off faster, as at most $2\Delta/(\tau\omega^2)$ (SM, Section~S2). Faster clearing therefore leaves more of the disconnection above $\omega_c$ where workload cannot imitate it. Together, \eqref{eq:d_lower} and this spectrum show that separation improves with \textit{(i)} a larger reduction relative to measurement noise, \textit{(ii)} faster clearing, \textit{(iii)} slower workload response at the POI, and \textit{(iv)} a wider recorder bandwidth.

The results above rely on the linear model~\eqref{eq:hypotheses}, the assumption that workload changes after disconnection produce no POI response they could not produce with all zones connected, and idealized bandwidths, and only $P_e$ depends on the Gaussian noise assumption. They also take the responses $\m b_N$ and $\m b_{\mathcal S}$ and the maps $\mathcal P_N$ and $\mathcal P_{\mathcal S}$ as known, which in practice requires a validated dynamic model of the facility. The implementation and the case studies do not use these assumptions. Even with the facility model known exactly, the margin is small in physical units. For a ten-zone facility of the form~\eqref{eq:zone}--\eqref{eq:hybrid} in which three zones disconnect over 16 or 50~$\ms$, the best workload-only match reproduces the POI current to within about 0.01~$\pu$ RMS, less than 3\% of the current removed and far below the prediction error of up to 22\% found in the case studies, so separating the mechanisms from POI data alone would require a facility model of that accuracy.

\section{Attribution With Workload Data}\label{sec:estimator}

When terminal measurements do not provide sufficient separation, workload data can be used to construct an all-connected response. The measured current is then compared with that response over a window, and the participation is estimated and its residual checked against an error tolerance.

\subsection{Estimating all-connected response}\label{subsec:network_ref}
The all-connected response $\m i^{(N)}$ in Section~\ref{subsec:participation} cannot be measured during an event because only the response of the equipment that actually remained connected is observed. It must be predicted by a model fitted on calibration records in which all zones are known to be connected ($\hat{\m g}_v$), which gives a predicted response $\hat{\m i}^{(N)}$. A natural choice is to use the measured POI voltage and workload histories as inputs. The predicted all-connected current is then
\[
\hat{\m i}^{(N)}_t=\hat{\m g}_v(\mathcal H_t^{v},\mathcal H_t^{w}).
\]
The difficulty is that the POI voltage is not independent of the facility current. In the quasi-static limit,~\eqref{eq:grid} gives
\begin{equation}\label{eq:network_inversion}
\m i_T=\m Z_g^{-1}(\m v_g-\m v),\qquad
\m Z_g=\begin{bmatrix}R_g&-X_g\\X_g&R_g\end{bmatrix},
\end{equation}
with $X_g=\omega_0L_g$. Thus, when the source voltage varies little, the measured POI voltage already contains information about the measured current through the grid relation. Therefore, if one fits a model to predict current using POI voltage, it can learn the network response rather than the all-connected response, while still achieving low calibration error.

This can be directly seen when $v_{g,q}(t)=\bar v_{g,q}$ and $i_q=0$. The $q$ component of~\eqref{eq:grid} simply becomes
\begin{equation}\label{eq:q_inversion}
i_d=\frac{\bar v_{g,q}-v_q}{X_g}.
\end{equation}
The derivation of~\eqref{eq:q_inversion} and the corresponding reconstruction error relation are given in SM, Section~S3. This is imposed by the network and does not depend on which equipment remains connected. A fitted model with sensitivity
$\partial\hat i_d/\partial v_q\approx-1/X_g$
can hence continue tracking the measured current after a disconnection. If it is used as the all-connected response, the predicted current falls with the measured current and the inferred participation can remain close to one even though equipment has disconnected. Comparing the fitted voltage sensitivity with $-1/X_g$ can therefore help identify a model that has learned the grid relation rather than the facility response.

To avoid using endogenous POI voltage as input, we first reconstruct the external source voltage from the POI data. With estimates $\hat R_g$ and $\hat L_g$, we get
\begin{equation}\label{eq:source_reconstruction}
\hat{\m v}_g=\m v+\hat R_g\m i_T+\hat L_g\dot{\m i}_T+\omega_0\hat L_g\m J\m i_T,
\end{equation}
where the derivative term is evaluated with filtering consistent with the recorder bandwidth. A separate model $\hat{\m g}_g$ is fitted on all-connected calibration records using reconstructed source voltage and workload histories, and the all-connected response is evaluated as
$$\hat{\m i}^{(N)}_t=\hat{\m g}_g(\mathcal H_t^{\hat v_g},\mathcal H_t^{w}).$$
In the noise-free model with the correct impedance and exact derivative, we get $\hat{\m v}_g=\m v_g$, so the exact POI voltage-current relation in~\eqref{eq:network_inversion} is no longer available to the predictor through its voltage input. With impedance errors $\Delta R_g=\hat R_g-R_g$ and $\Delta L_g=\hat L_g-L_g$, the reconstructed voltage is given by
$\hat{\m v}_g = \m v_g +\Delta R_g\m i_T +\Delta L_g\dot{\m i}_T +\omega_0\Delta L_g\m J\m i_T.$
The residual dependence on measured current is therefore proportional to the impedance error. This error, together with filtering and measurement errors, is included in the prediction error. We note here that while calibration records containing independent source voltage variation (such as phase changes) can help distinguish the grid relation from the facility response, they do not by themselves guarantee that the fitted model represents the facility response.

Because a phase jump moves $\hat v_{g,q}$, to which a fitted correction can be sensitive, the correction is fitted and evaluated in a frame aligned with $\hat{\m v}_g$ and rotated back (SM, Section~S3); restricting its voltage inputs to $\|\hat{\m v}_g\|$ also removes the path of~\eqref{eq:q_inversion}. Its inputs and ridge parameter are chosen by leave-one-out on the calibration records.

Reconstructing the source voltage requires a validated grid impedance, which can be supplied from interconnection or short-circuit data. Identifying it from records can be suitable when available excitation supports identifying both $R_g$ and $L_g$. We note here that a fault also changes the impedance seen from the POI (both during the fault and after clearing) and by~\eqref{eq:source_reconstruction} this change enters the reconstructed voltage in proportion to the facility current. The impedance must therefore match the network state over the attribution window. In post-event analysis, the fault location and clearing times are known from protection records, so the impedance for each interval can be computed from the network model. In the case studies the selected prediction did not reproduce the grid relation even with measured POI voltage, and impedance errors of $\pm10\%$, or different errors during and after the sag changed no classification (see Section~\ref{subsec:cs_limits}); the source voltage still gave smaller errors and excludes~\eqref{eq:network_inversion} by construction rather than through the fit.

\subsection{Participation range and feasibility}\label{subsec:certificate}

Having defined the prediction, collect the measured and predicted currents over a window $\mathcal W$ as $\m i_{\mathcal W},\hat{\m i}^{(N)}_{\mathcal W}\in\RR^{2n}$. Then for a candidate participation value $a\in[0,1]$, define the current residual as
\begin{equation}\label{eq:one_factor}
\m r_{\mathcal W}(a) = \m i_{\mathcal W} -a\,\hat{\m i}^{(N)}_{\mathcal W}.
\end{equation}
The candidate $a$ is considered consistent with the data when
\[
\|\m r_{\mathcal W}(a)\|_2\le\varepsilon\sqrt{2n},
\]
where $\varepsilon$ bounds the allowed per-component root mean square discrepancy. The admissible participation set is therefore given by
\[
\mathcal A(\varepsilon) = \big\{a\in[0,1]\mid\|\m i_{\mathcal W}-a\hat{\m i}^{(N)}_{\mathcal W}\|_2 \le\varepsilon\sqrt{2n}\big\}.
\]
Let
\[
A=\hat{\m i}^{(N)\top}_{\mathcal W}\hat{\m i}^{(N)}_{\mathcal W},\qquad B=\hat{\m i}^{(N)\top}_{\mathcal W}\m i_{\mathcal W},
\]
\[
C=\m i_{\mathcal W}^{\top}\m i_{\mathcal W}-2n\varepsilon^2,\qquad D=B^2-AC.
\]

The participation values consistent with the measured POI current are then
\begin{equation}\small\label{eq:interval}
\mathcal A(\varepsilon)=
\begin{cases}
\left[\dfrac{B-\sqrt D}{A},\dfrac{B+\sqrt D}{A}\right]\cap[0,1], & A>0,\ D\ge0,\\[1ex]
[0,1], & A=0,\ \|\m i_{\mathcal W}\|_2\le\varepsilon\sqrt{2n},\\
\emptyset, & \text{otherwise},
\end{cases}
\end{equation}
which follows from writing the admissibility condition as a quadratic inequality in $a$ (see SM, Section~S4). A proper interval would give a participation range supported by the data, while $[0,1]$ is uninformative and $\emptyset$ means that no participation value is consistent with the error tolerance we chose. The set in~\eqref{eq:interval} is the exact solution of the residual test for a prescribed tolerance. It is a consistency set, and its inclusion of the true participation depends on the true residual remaining within that tolerance. In particular, if the residual evaluated at the ideal electrical participation $a_e(\mathcal W)$ in~\eqref{eq:participation} satisfies the stated condition, then $a_e(\mathcal W)\in\mathcal A(\varepsilon)$. For $A>0$, the unconstrained least-squares ratio is $a_{\rm raw}=B/A$, and the corresponding estimate is then simply $\hat a=\min\{1,\max\{0,a_{\rm raw}\}\}$.

When the fit is well within the tolerance, the interval has half-width of approximately $\varepsilon/\bar i$, where $\bar i=\|\hat{\m i}^{(N)}_{\mathcal W}\|_2/\sqrt{2n}$ is the root mean square of the predicted current. The set therefore widens with the tolerance and narrows as the predicted current grows, so an overestimated prediction can make the set look more precise than it actually is.

The calibration records must have a known connected set and cover the conditions for which attribution is claimed. Commissioning tests (which the NERC Level~3 alert calls for~\cite{nerc_l3_2026}) can supply them, as can records of disturbances that the facility rode through, confirmed by its event logs, or simulations of a validated facility model. We define $\varepsilon$ as the largest per-component root mean square residual at $a=1$ over the windows of calibration records not used to fit the prediction (leave-one-out when records are few), separately for each window type and only over windows that satisfy~\eqref{eq:nose}. $\mathcal A(\varepsilon)$ is then a consistency check, not an uncertainty statement: it contains the participation values for which the measured current is a scaled copy of the prediction to within the calibration error, and it is empty when the current contains a component the prediction does not scale, such as filter current after a cessation or units reconnecting in the window. The attribution is then $\hat a$, and its uncertainty is stated from validation and, where the set is non-empty, from its half-width. Attribution is withheld when the window lies outside the calibrated range, that is, when the minimum of $\|\hat{\m v}_g\|$ over the window falls more than $\delta_v$ below its smallest value in the calibration records, or the workload data leave their calibrated range by more than $\delta_w$. The window must be placed using quantities available in the record, since the time of any protection action is unknown.

In power, with $\hat P^{(N)}$ the window mean of $\m v_t^\top\hat{\m i}^{(N)}_t$ (measured POI voltage) and $P_{\rm pre}$ the mean POI power in the 100~$\ms$ before the disturbance, the load lost to disconnection is $(1-\hat a)\hat P^{(N)}$ and the load change from the workload $P_{\rm pre}-\hat P^{(N)}$. The two terms add to $P_{\rm pre}-\hat a\hat P^{(N)}$, which equals the measured power drop when the measured current is a scaled copy of the prediction, otherwise the residual carries the remainder. Both use the measured POI voltage and so neglect the voltage change caused by the disconnection itself.

The prediction $\hat{\m i}^{(N)}_t$ must also be physically plausible throughout the window. A deep enough source voltage sag leaves no operating point at which the facility can draw its demanded power. For a facility drawing $P_T>0$ at unity power factor behind the grid impedance in~\eqref{eq:network_inversion}, with $|Z_g|=\sqrt{R_g^2+X_g^2}$, such an operating point exists if and only if
\begin{equation}\label{eq:nose}
\|\m v_g\|\ge v_{\rm nose}=\sqrt{2P_T\left(|Z_g|+R_g\right)},
\end{equation}
the minimum source voltage that can deliver $P_T$ through $Z_g$ (see SM, Section~S5, which also gives the condition when the converter current is in phase with the source voltage). Below $v_{\rm nose}$, the prediction is not meaningful and attribution should be withheld. Converter current limits and control modes can move the actual boundary, and must then be represented in the facility model.

When a facility has several protection zones with different dynamics, a single participation value is only an approximation. A closed-loop facility model can instead be evaluated for each candidate connected set, and with zone-level workload data the sets consistent with the measured current identify the zones that disconnected. With aggregate workload data the workload of a candidate set must be allocated from the total, which introduces an error that vanishes only when connected and disconnected zones fluctuate alike per unit nominal power. The formulation is in the supplement (Section~S6).

\section{Data Requirements for Attribution}\label{sec:requirements}
The NERC Level~3 alert calls for fault recording as well as protection and facility modeling information~\cite{nerc_l3_2026}. The attribution results in this work add the following requirements for each kind of information.
\begin{itemize}[leftmargin=*]
\item \textit{POI voltage and current} separate the two mechanisms on their own only through the part of a disconnection that is faster than any workload change~\eqref{eq:d_lower}, so the recorder must retain the fast transient, with a bandwidth above the workload response bandwidth, and phasor-rate data are not enough during the disturbance (Section~\ref{subsec:cs_limits}).
\item \textit{Aggregate workload data} allow the all-connected response, and the share of it that remains at the POI, to be estimated. They must represent the computational demand $P_k(t)$ over the attribution window, including after a disconnection or transfer to backup supply. Power measured at the facility input does not, because it drops whenever equipment disconnects, so a prediction based on it drops with the measured current and the estimated participation stays close to one. Power at the UPS output does, since it keeps supplying the computing equipment during a transfer to battery. The data must also be fast and aligned in time with the POI record (Section~\ref{subsec:cs_limits}).
\item \textit{Calibration records} must cover the voltage and workload range of the event, because the prediction is used only within that range, and records of disturbances that the facility rode through do not reach the sag depths at which it trips (Section~\ref{subsec:cs_pscad}).
\item \textit{Zone-level workload data} allow candidate connected sets to be evaluated, which identifies the zones that disconnected. With only aggregate data, the workload of a candidate set must be allocated from the total, which introduces an allocation error (SM, Section~S6).
\item \textit{A validated grid impedance} is needed to reconstruct the source voltage, which keeps~\eqref{eq:network_inversion} out of the prediction by construction. The error in~\eqref{eq:source_reconstruction} is proportional to the impedance error, and the impedance must match the network state during and after the fault, although impedance errors of $\pm10\%$, or different errors during and after the sag, changed no classification in the tests (Section~\ref{subsec:cs_limits}).
\end{itemize}
\section{Case Studies}\label{sec:case_studies}

\subsection{Test systems and procedure}\label{subsec:cs_setup}

Table~\ref{tab:cs_testbeds} summarizes the three PSCAD test systems, two derived from the PNNL Data Center Model Library~\cite{ross2026dml} and one from the TAMU/ERCOT crypto-miner model~\cite{wang2025crypto}, and the versions of them with separate zone settings. Before the disturbance, the POI active power of DC2 and DC3 is about 803 and 813~MW, which includes support equipment and conversion losses. For the PNNL cases, we removed pump and fan branches to satisfy the 200 electrical node limit of the PSCAD educational license. In each test system, the upstream network is a controlled Th\'evenin source behind a known $R_g$--$L_g$ impedance, and the workload is specified independently of which equipment is connected. The prediction model of each test system combines a quasi-static constant power response driven by the filtered workload with a linear correction fitted by ridge regression on the reconstructed source voltage and workload in the frame of Section~\ref{subsec:network_ref}, using their filtered histories at 2, 10, and 50~$\ms$. Leave-one-out on the calibration records chose magnitude-only voltage inputs and a ridge parameter 0.1 to 1000 times the original in 10 of 12 calibrations. Ridge factors above one were added to the candidates after a first round before any evaluation event was scored with them.

\input{tables/Table_CS1_testbeds.tex}

Each event is simulated twice with the same source voltage, workload, initial conditions, and solver settings, once with protection acting normally and once with the disconnection suppressed. For an evaluation event, the all-connected run only gives the reference participation $a^\star$ from~\eqref{eq:participation} and checks the prediction. The estimator receives the POI voltage and current at 20~kHz without added noise unless stated otherwise, workload data at 1~kHz with $10^{-3}$~$\pu$ quantization, and the grid impedance. It does not receive the all-connected run of an evaluation event, the source voltage, or the protection channels. The current derivative in~\eqref{eq:source_reconstruction} is filtered with a 1~$\ms$ exponential filter.

We use 21 calibration and 37 evaluation events with workload changes, protection disconnections, or both. In most of them the facility rides through, or all of its converters stop drawing grid current at once (called cessation below), so we added 60 evaluation events on versions of the three models whose zones have different undervoltage settings, in which some zones trip and others ride through. A second UPS zone in DC2 or a second PFC rack in DC3 exceeds the node limit, so DC2 uses its UPS (91\% of the pre-event power) and support (9\%) branches as zones, and DC3 gives its three single-phase PFC groups different settings, so that they cease phase by phase. TAMU is split into three equal groups of miner units. The zone thresholds are 0.80, 0.75, and 0.70~$\pu$ with delays of 50, 80, and 120~$\ms$; DC2 uses 0.70~$\pu$/50~$\ms$ for the UPS and 0.80~$\pu$/120~$\ms$ for the support branch, and the DC3 support branch uses 0.75~$\pu$/50~$\ms$. The events combine three sag depths, two durations, and workload steps of $-20\%$, 0, and $+15\%$, and include two events per test system with a $20^\circ$ phase jump lasting until recovery. The DC2 and DC3 zone settings change the all-connected response to deep sags, so their calibration runs were regenerated. After a first scoring showed that the jump distorted the prediction, three ride-through events with a $20^\circ$ jump were added to both calibrations of each version with zones. With them, the predicted to all-connected current ratio of the phase-jump events lies within the range of the other events in five of six cases (test system and window) and exceeds it by 0.0007 in the sixth (TAMU after recovery), which both voltage components as inputs would have avoided.

Windows are placed from the record alone. The onset $t_{\rm on}$ is the first time at which $\|\hat{\m v}_g\|$ stays below $0.90\,\bar v_{\rm pre}$ for 2~$\ms$, or the workload data depart from their pre-event mean by more than 2\% of the rating for 2~$\ms$, where $\bar v_{\rm pre}$ is the mean of $\|\hat{\m v}_g\|$ over the preceding 100~$\ms$. The calibrated range uses $\delta_v=0.02~\pu$ and $\delta_w=2\%$ of the rating. The disturbance window $\mathcal W_{\rm dist}$ spans the 300~$\ms$ after $t_{\rm on}$. The recovery time $t_{\rm rec}$ is the first later time at which $\|\hat{\m v}_g\|$ stays at or above $0.90\,\bar v_{\rm pre}$ for 20~$\ms$, and the post-recovery window $\mathcal W_{\rm post}$ spans $[t_{\rm rec}+50,\,t_{\rm rec}+250]$~$\ms$.

We calibrate the prediction in two ways. The ride-through calibration uses only records of disturbances the facility rode through, which a facility could supply from its own event history: six DC2, four DC3, and four TAMU events, and eight, seven, and eight for the versions with zones, including the phase-jump events. The model calibration fits the prediction on all-connected simulations of six DC2, three DC3, and three TAMU events (nine, six, and six for the versions with zones), which assumes a validated facility model. In both, $\varepsilon$ follows Section~\ref{subsec:certificate} (Table~\ref{tab:cs_results}). We report the ride-through calibration on $\mathcal W_{\rm post}$ and the model calibration on $\mathcal W_{\rm dist}$.

A window is admitted if the reconstructed source voltage magnitude, filtered with a 10~$\ms$ exponential filter, satisfies~\eqref{eq:nose} and the window is inside the calibrated range. We added the workload part of this range check after we had seen the partial-trip results for TAMU, where three of the four estimates more than 0.10 above the reference had a workload step to 115\% of the rating, above the 110\% of the calibration records. It withholds the six DC3 and six TAMU events with this step and four DC3 and TAMU events of the first 37 whose workload falls to 60--65\%, below the 80\% of their calibration records. These four played no part in motivating the check and would be classified correctly without it. A test on the calibration records did not support it either, a prediction fitted on workloads up to 100\% stayed within $\varepsilon$ at 110\% in all four cases, and one fitted from 90\% upward exceeded $\varepsilon$ at the lowest workload in five of nine. The range is therefore a stated validity range and Section~\ref{subsec:cs_pscad} reports the withheld events. Two further TAMU events with a shallow sag and no workload change open no window ($\|\hat{\m v}_g\|\ge0.90\,\bar v_{\rm pre}$). Every other rule was fixed before any evaluation event was scored with it. An event is a loss if $a^\star<0.95$ on the window and partial if $0.05<a^\star<0.95$; it is classified as a loss if $\hat a<0.95$, so a false loss is a loss-free event classified as a loss and a missed loss a loss event classified as loss-free.

\subsection{Value of workload data}\label{subsec:cs_information}

\begin{figure}[t]
\centering
\includegraphics[width=0.86\columnwidth]{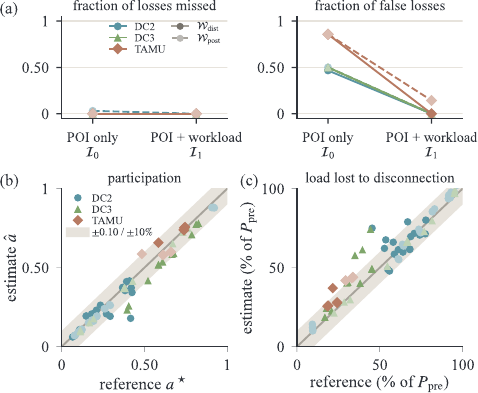}
\caption{Both sets of events. (a) Fraction of admitted loss events missed and of admitted loss-free events classified as losses, with POI data alone and with workload data. (b) Estimated versus reference participation for every admitted event and window with $0.05<a^\star<0.95$. (c) Estimated versus reference load lost to disconnection for the same windows, in percent of the pre-event power.}
\label{fig:cs_information}
\end{figure}

Fig.~\ref{fig:cs_information}(a) compares $\mathcal I_0$ and $\mathcal I_1$ in~\eqref{eq:information_sets}. For $\mathcal I_0$ the workload input of the same prediction is held at its pre-event value, with the windows and admission of $\mathcal I_1$. With POI data alone, 47--86\% of the admitted loss-free events are classified as losses, all with a workload reduction of 10--40\%, which produces the same POI response as a disconnection (Section~\ref{subsec:static}). With workload data, none is during the disturbance and one of seven TAMU events is after recovery. POI data alone miss one DC2 loss after recovery and workload data none. Without the workload part of the range check, POI data alone missed one of 15 TAMU losses after recovery and workload data none, and workload data classified one of nine loss-free TAMU events as a loss.

\subsection{Results on PSCAD models}\label{subsec:cs_pscad}

\begin{figure}[t]
\centering
\includegraphics[width=0.9\columnwidth]{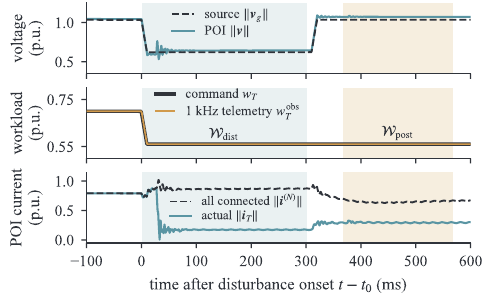}
\caption{DC2 event with a source voltage sag and a workload reduction. $\mathcal W_{\rm dist}$ and $\mathcal W_{\rm post}$ are shaded.}
\label{fig:cs_event}
\end{figure}

\input{tables/Table_CS2_results.tex}

In the DC2 event of Fig.~\ref{fig:cs_event}, the POI current would fall even if nothing disconnected, as the all-connected run shows, but it falls further because protection removes part of the facility's current, and the estimator separates the two effects using the prediction. In the first 37 events (Table~\ref{tab:cs_results}) the ride-through calibration after recovery gives an estimate for 30 events; three have no window and four are outside the workload range of the calibration. None of the seven admitted loss-free events is classified as a loss and none of the 23 loss events is missed, including 11 events below the feasibility limit during the sag, four of them in TAMU, whose breakers stay open. During the disturbance the ride-through calibration admits no loss event, because ride-through records do not reach the sag depths at which a facility trips. With the model calibration the eight admitted DC2 loss events have errors of at most 0.045, and only one DC3 event and no TAMU event are admitted. The DC3 event has an error of 0.104 against a tolerance of 0.52~$\pu$, so the DC3 model calibration is almost uninformative during the sag. For 19 of the 23 loss events after recovery $\mathcal A(\varepsilon)$ is empty, and in 15 of them at least 95\% of the residual $\m i_{\mathcal W}-\hat a\,\hat{\m i}^{(N)}_{\mathcal W}$ is reactive. The check thus flags the current of input filters that stay energized after a cessation (Section~\ref{subsec:static}), and the attribution is the estimate $\hat a$.

In the 60 partial-trip events the ride-through calibration after recovery is within 0.10 of the reference for all 27 admitted partial events of DC2 and DC3 (maximum error 0.044, $a^\star$ from 0.08 to 0.92), with no false or missed loss (Fig.~\ref{fig:cs_information}(b)). They include UPS cessation, support-branch trips, and cessation on one, two, or three phases. For TAMU none of the five admitted partial events is missed (largest error 0.102), but one loss-free event is classified as a loss ($\hat a=0.948$). The set excludes the reference in one DC2 and three TAMU events, by 0.006 to 0.067. The withheld TAMU partial events with the 115\% step would be classified correctly, but three have errors of 0.12 to 0.39 because the all-connected miners then draw more current than predicted (ratio 0.50--0.92); the DC3 events with this step have errors of at most 0.021.

During the disturbance the ride-through calibration admits no partial event. With the model calibration, 24 of 28 admitted partial events are within 0.10. The predicted current of the phase-jump events is 0.93 to 1.11 times the all-connected current in DC2 and DC3 (0.99 to 1.01 in TAMU), and none of the five partial phase-jump events is outside 0.10. The four errors above 0.10 are DC2 and DC3 events in which every zone trips during the window, underestimated by 0.10 to 0.24, and no loss-free event is classified as a loss. The withheld events with the 115\% step would have errors of up to 0.13 (TAMU) and 0.17 (DC3), all classified correctly. After recovery the ride-through calibration admits the phase-jump events and classifies them correctly, with errors within 0.003 of those of the same events without the jump.

In percent of the pre-event power (Table~\ref{tab:cs_results}, Fig.~\ref{fig:cs_information}(c)), the error of the load lost to disconnection after recovery is at most 6.8\% for every admitted DC2 and DC3 event and that of the workload change at most 10.2\%; during the disturbance the model calibration reaches 30\% where the participation is underestimated. For TAMU the load lost to disconnection is off by 10.1 to 12.5\% in all five partial events after recovery and by 97\% on average in the first 37 events, because the prediction does not represent the all-connected current after the sag (Section~\ref{subsec:cs_recovery}).

\subsection{Feasibility limit and sensitivity}\label{subsec:cs_limits}

For the admitted events the ratio of predicted to simulated all-connected current (SM, Fig.~S1) is 0.89--1.22 during the disturbance, and after recovery 1.00--1.05 for DC2 and DC3 and 0.53--3.68 for TAMU (Section~\ref{subsec:cs_recovery}). Below the feasibility limit the constant power prediction demands more current as the source voltage falls, and eight of the 13 events withheld by~\eqref{eq:nose} have predicted currents 2.0 to 31.7 times the simulated value. The PNNL cases operate at about 0.97 leading power factor while~\eqref{eq:nose} assumes unity power factor, so here the condition is only an approximate limit. Using measured POI voltage instead of the reconstructed source voltage raises the mean absolute error after recovery from 0.012 to 0.022 and misses none of the 23 loss events, so with the inputs and regularization selected on calibration records the failure of Section~\ref{subsec:network_ref} did not occur.

\begin{figure}[t]
\centering
\includegraphics[width=0.9\columnwidth]{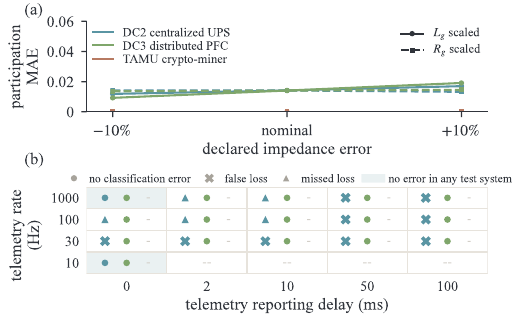}
\caption{Effect of (a) grid impedance errors, with resistance and inductance perturbed separately (ride-through calibration, $\mathcal W_{\rm post}$), and (b) workload data rate and reporting delay (model calibration, $\mathcal W_{\rm dist}$), first 37 events. Dashes mark untested cells and test systems without an admitted event.}
\label{fig:cs_requirements}
\end{figure}

A $\pm10\%$ error in resistance or inductance changes no classification (Fig.~\ref{fig:cs_requirements}(a)), and the DC2 mean error after recovery stays between 0.012 and 0.017. With an impedance 30\% too low during the sag and 10\% too high after it, or the reverse, no classification changes in either calibration, and the DC2 mean error after recovery is at most 0.015.

During the disturbance, workload data at 100~$\Hz$ or more with delays up to 10~$\ms$ change the estimate by at most 0.014, and the largest change over the tested rates and delays is 0.083 (30~$\Hz$, 100~$\ms$; Fig.~\ref{fig:cs_requirements}(b)). Because the classification threshold equals that of the label, 14 of the 16 tested combinations change a classification, all for four DC2 events near 0.95: one whose reference sits at the boundary ($a^\star=0.950$) changes label as the window moves, and three loss-free events with estimates of 0.98 to 0.99 fall below it. After recovery no tested combination of rate (10--1000~$\Hz$) and delay (up to 100~$\ms$) changes any classification, and the estimate changes by at most 0.026. In post-event analysis the delay is a time offset, so accurate timestamps matter; power quality meters report half-cycle RMS values~\cite{iec61000430} synchronized to within 1~$\ms$ by the precision time protocol~\cite{ieee1588}.

With white sensor noise of $\sigma_v=0.0027$ and $\sigma_i=0.0024$~$\pu$ per 20~kHz sample (three realizations, recalibrated), no classification changes in the first 37 events, and after recovery in the partial-trip events only the TAMU loss-free event at $\hat a=0.948$ changes, in two realizations. The post-recovery error rises by at most 0.003, and the error of the one admitted DC3 event during the disturbance is 0.103 in every realization. Averaging to 60 frames per second changes no post-recovery classification in the first 37 events, classifies one more TAMU loss-free event as a loss after recovery, and raises the TAMU partial-trip error from 0.042 to 0.050. During the disturbance it classifies six of 11 admitted loss-free events of the first 37 and ten of 20 of the partial-trip events as losses, eight of them TAMU.

\subsection{Participation after recovery}\label{subsec:cs_recovery}

\begin{figure}[t]
\centering
\includegraphics[width=0.9\columnwidth]{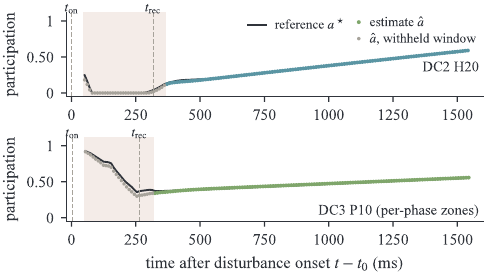}
\caption{Participation over sliding 100~$\ms$ windows (ride-through calibration) for the event of Fig.~\ref{fig:cs_event} and a DC3 partial trip in which two phase groups and the support branch ceased. Shaded: withheld windows. $\mathcal A(\varepsilon)$ is empty in every window, so no range is drawn.}
\label{fig:cs_timeline}
\end{figure}

In DC2 and DC3 the units that ceased start to reconnect 56 to 63~$\ms$ after $t_{\rm rec}$, so the participation rises during $\mathcal W_{\rm post}$ (Fig.~\ref{fig:cs_timeline}), from 0.04--0.30 at $t_{\rm rec}$ to 0.07--0.34 at $t_{\rm rec}+50$~$\ms$ and 0.18--0.41 at $t_{\rm rec}+500$~$\ms$, and it is still rising at the end of the records (0.30--0.73). The estimate follows with a lag: for the event of Fig.~\ref{fig:cs_event}, $a^\star$ and $\hat a$ are 0.155 and 0.143 at $t_{\rm rec}+20$~$\ms$ and 0.205 and 0.207 at $t_{\rm rec}+200$~$\ms$, and $\mathcal A(\varepsilon)$ is empty in every window of both events. The post-recovery number therefore describes a facility whose units are reconnecting, not the share that disconnected during the fault, and it depends on where the window is placed. In TAMU the breakers stay open, so a loss event keeps zero participation, but the all-connected run draws 2.50 to 2.61 times its pre-event current after the sag and more than 1.5 times for 0.12 to 0.60~s, which the prediction does not represent. The TAMU estimates are exact only because the measured current is about zero; when part of the facility stays connected, the same mismatch biases the estimate, as in the TAMU partial events, and causes the TAMU load errors.

\section{Conclusion}\label{sec:conclusion}

This work addressed the attribution of large computational load reductions to workload changes or protection disconnections. In a quasi-static constant power model the two mechanisms produce identical POI voltage and current. Their transient responses separate only through the part of a disconnection faster than any workload change, which requires a facility model far more accurate than calibration achieves, so attribution needs workload data that represent computational demand independently of the grid connection, a grid impedance matched to the network state, and calibration records that cover the event. A prediction fitted to measured POI voltage can learn the grid relation and hide a disconnection; with inputs and regularization chosen on calibration records it did not in these tests, and the reconstructed source voltage gave smaller errors.

With records of disturbances the facility rode through, the estimator needs no facility model. After voltage recovery it classified every admitted event of the two data center architectures correctly, including partial disconnections, and estimated the load lost to disconnection within 7\% of the pre-event power. For the crypto-miner facility it classified one of seven loss-free events as a loss, and its load estimates were off by 10 to 13\% because the prediction does not represent the current surge after the sag. The post-recovery estimate describes a facility whose units may already be reconnecting. During the disturbance, ride-through records do not reach the sag depths at which a facility trips, so attribution there needs a validated facility model, and even then one in seven partial events was off by more than 0.10.

The results rest on EMT models with a controlled Th\'evenin source, a known impedance, and no field data, and on a workload signal that remains valid after disconnection. The consistency check $\mathcal A(\varepsilon)$ is rarely informative after a cessation so the uncertainty of the estimate rests on its validation. The workload range of the calibration is a stated validity range that a test on the calibration records did not support. Auxiliary mechanical loads were removed from the models; their contactor dropout would appear as a disconnection, but no workload signal exists for them. Field validation with synchronized POI and workload measurements is the next step.

\section*{Acknowledgment}
The authors used an AI language model (Claude, Anthropic) to assist in editing the text and in implementing the proposed methods. The authors reviewed all content and are responsible for it.

\balance
\bibliographystyle{IEEEtran}
\bibliography{refs}

\setcounter{section}{0}
\renewcommand{\thesection}{S\arabic{section}}
\setcounter{equation}{0}
\renewcommand{\theequation}{S\arabic{equation}}
\renewcommand{\theHsection}{S\arabic{section}}
\renewcommand{\theHequation}{S\arabic{equation}}
\setcounter{figure}{0}
\renewcommand{\thefigure}{S\arabic{figure}}
\renewcommand{\theHfigure}{S\arabic{figure}}
\setcounter{table}{0}

\begin{center}
{\large\bfseries Supplementary Material}\\[3pt]
{\bfseries Data Center Load Reductions: Attribution, Measurement Limits, and Estimation.}\\[2pt]
Abdallah Alalem Albustami, Ahmad F.~Taha, Mohammad Alwahdanee, and Zuhaib Shakeel Muhammad
\end{center}
\vspace{2pt}

\noindent This supplement collects the derivations of the relations used in the paper. Equation numbers of the form~(k) refer to the main paper and numbers of the form~(Sk) to this supplement. Notation follows the main paper. All quantities are in a power-invariant synchronous $dq$ frame, $\m J$ is the rotation by $90^\circ$, the facility is connected through a Th\'evenin source $\m v_g$ and impedance $(R_g,L_g)$ with $X_g=\omega_0L_g$ and $|Z_g|=\sqrt{R_g^2+X_g^2}$, and the quasi-static form of the grid equation~\eqref{eq:grid} is $\m v=\m v_g-\m Z_g\m i_T$ with $\m Z_g$ as in~\eqref{eq:network_inversion}.

\section{Participation and Connected Power}\label{sup:ae}

This section derives~\eqref{eq:ae_capacity}. The setting is quasi-static constant power operation at unity power factor, zone powers $P_k=\gamma P_k^0$ for a common loading factor $\gamma>0$, negligible feeder drops, and POI voltage magnitudes $\bar v^{(N)}$ and $\bar v^{(\mathcal S)}$ that are constant over the window in the all-connected and partially connected cases.

With negligible feeder drops, all connected zones see the same POI voltage. The all-connected facility draws $P_T=\gamma\sum_kP_k^0$ and the partially connected facility draws $c(\mathcal S)P_T$, so the current magnitudes are $P_T/\bar v^{(N)}$ and $c(\mathcal S)P_T/\bar v^{(\mathcal S)}$. If the phase difference between the two POI voltages is neglected, the two current vectors are collinear at every sample, and the unconstrained least-squares ratio of the measured current onto the all-connected current is
\begin{equation}\label{eqS:ratio}\small
\frac{\langle\m i_T,\m i^{(N)}\rangle_{\mathcal W}}{\langle\m i^{(N)},\m i^{(N)}\rangle_{\mathcal W}}
=\frac{n\,\big(c(\mathcal S)P_T/\bar v^{(\mathcal S)}\big)\big(P_T/\bar v^{(N)}\big)}{n\,\big(P_T/\bar v^{(N)}\big)^2}
=c(\mathcal S)\,\frac{\bar v^{(N)}}{\bar v^{(\mathcal S)}}.
\end{equation}
The inequality $\bar v^{(N)}\le\bar v^{(\mathcal S)}$ follows from the voltage ordering on the high-voltage branch. Let $x=V^2$ with $V=\|\m v\|$ and let $P$ be the connected constant power. At unity power factor $\m i_T=(P/V^2)\m v$, and the quasi-static grid equation gives
\begin{equation}\label{eqS:vg_norm}
\|\m v_g\|^2=x+2R_gP+\frac{|Z_g|^2P^2}{x}.
\end{equation}
Holding $\|\m v_g\|$ fixed and differentiating implicitly with respect to $P$,
\begin{equation}\label{eqS:dxdP}
\frac{\dd x}{\dd P}=\frac{2x\,(P|Z_g|^2+R_gx)}{P^2|Z_g|^2-x^2}.
\end{equation}
On the high-voltage branch $x>|Z_g|P$, the denominator is negative and the numerator is positive, so $\dd x/\dd P<0$. Reducing the connected power therefore raises the POI voltage, which gives $\bar v^{(\mathcal S)}\ge\bar v^{(N)}$ and hence the ratio in~\eqref{eqS:ratio} lies in $[0,c(\mathcal S)]$. Since the ratio lies in $[0,1]$, the constrained definition~\eqref{eq:participation} returns it unchanged, which is~\eqref{eq:ae_capacity}.

\section{Dynamic Separation}\label{sup:sep}

This section derives~\eqref{eq:separation}, the error probability $P_e=Q(d/2)$, and the bound~\eqref{eq:d_lower}. The setting is the two hypotheses~\eqref{eq:hypotheses} on a window of $n$ samples, with $\m y\in\RR^{4n}$ the whitened POI voltage and current deviations, $\m e\sim\mathcal N(\m 0,\m I)$, $\mathcal U$ a linear subspace of admissible workload deviations, and $\mathcal P_{\mathcal S}\mathcal U\subseteq\mathcal P_N\mathcal U$.

\subsection{Reduction of the response-set distance}
Let $\tilde{\m y}=\m y-\m b_N$ and $\m s_{\mathcal S}=\m b_{\mathcal S}-\m b_N$. Then $\tilde{\m y}=\mathcal P_N\m p+\m e$ under $\mathcal H_0$ and $\tilde{\m y}=\m s_{\mathcal S}+\mathcal P_{\mathcal S}\m p'+\m e$ under $\mathcal H_1$. The distance~\eqref{eq:general_separation} is the infimum over $\m p,\m p'\in\mathcal U$ of $\|\m s_{\mathcal S}-(\mathcal P_N\m p-\mathcal P_{\mathcal S}\m p')\|_2$. Since $\mathcal U$ is a subspace, $\mathcal P_N\mathcal U$ is a subspace, and $\mathcal P_{\mathcal S}\mathcal U\subseteq\mathcal P_N\mathcal U$ implies that every difference $\mathcal P_N\m p-\mathcal P_{\mathcal S}\m p'$ lies in $\mathcal P_N\mathcal U$. Conversely, taking $\m p'=0$ shows that every element of $\mathcal P_N\mathcal U$ is such a difference. The set of differences is therefore exactly $\mathcal P_N\mathcal U$, and the infimum is the distance from $\m s_{\mathcal S}$ to that subspace,
\begin{equation}\label{eqS:dist}
d_{\rm gen}(\mathcal S)=\dist(\m s_{\mathcal S},\mathcal P_N\mathcal U)=\|(\m I-\m\Pi)\m s_{\mathcal S}\|_2,
\end{equation}
with $\m\Pi$ the orthogonal projector onto $\mathcal P_N\mathcal U$. This is~\eqref{eq:separation}.

\subsection{Error probability of the projected test}
Let $\m\Pi^\perp=\m I-\m\Pi$. Because the workload term lies in $\mathcal P_N\mathcal U$ under both hypotheses, projecting removes it,
\begin{equation}\label{eqS:proj}
\m\Pi^\perp\tilde{\m y}=\m\Pi^\perp\m e\ \ (\mathcal H_0),\qquad
\m\Pi^\perp\tilde{\m y}=\m\Pi^\perp\m s_{\mathcal S}+\m\Pi^\perp\m e\ \ (\mathcal H_1).
\end{equation}
On the range of $\m\Pi^\perp$ the noise has identity covariance and the two means differ by $\m\Pi^\perp\m s_{\mathcal S}$, whose norm is $d(\mathcal S)$. The equal-prior likelihood-ratio test between two Gaussians with equal covariance compares $\m s_{\mathcal S}^\top\m\Pi^\perp\tilde{\m y}$ with $\tfrac12\|\m\Pi^\perp\m s_{\mathcal S}\|_2^2$ and has error probability $Q(d(\mathcal S)/2)$~\cite{kay1998}. Setting $Q(d/2)\le\alpha$ gives $d\ge2Q^{-1}(\alpha)$ for $0<\alpha<1/2$.

\subsection{Current-only lower bound}
Let the current-measurement noise be white with covariance $\sigma_i^2\m I$ before whitening and independent of the voltage-measurement noise, so that whitening acts on the current block by the scalar $1/\sigma_i$. Let $\bar{\mathcal P}_i$ and $\bar{\m s}_{\mathcal S,i}$ be the current blocks of $\mathcal P_N$ and $\m s_{\mathcal S}$ before whitening, let $\m\Pi_i$ be the orthogonal projector onto $\bar{\mathcal P}_i\mathcal U$, and define $d_i(\mathcal S)=\|(\m I-\m\Pi_i)\bar{\m s}_{\mathcal S,i}\|_2/\sigma_i$. From~\eqref{eqS:dist}, $d(\mathcal S)$ is the minimum over $\m p\in\mathcal U$ of the norm of a whitened vector whose current block is $(\bar{\m s}_{\mathcal S,i}-\bar{\mathcal P}_i\m p)/\sigma_i$. The norm of a vector is at least the norm of any one of its blocks, and the minimum of the block norm over $\m p$ is $d_i(\mathcal S)$. Hence $d(\mathcal S)\ge d_i(\mathcal S)$.

Suppose $\mathcal U$ contains finite-window DFT components only below $\omega_u$, the workload-to-current path is ideal low-pass with bandwidth $\omega_p$, and the recorder passband is flat to $\omega_r>\omega_c=\min\{\omega_u,\omega_p\}$. Then $\bar{\mathcal P}_i\mathcal U$ lies in the span of the DFT components of the window below $\omega_c$. The DFT components form an orthogonal basis, so the distance from $\bar{\m s}_{\mathcal S,i}$ to any subspace of that span is at least the norm of the recorded component of $\bar{\m s}_{\mathcal S,i}$ between $\omega_c$ and $\omega_r$, which is $\Delta\m i_{\rm fast}$ in~\eqref{eq:d_lower}. Dividing by $\sigma_i$ gives $d_i(\mathcal S)\ge\|\Delta\m i_{\rm fast}\|_2/\sigma_i$, which together with $d(\mathcal S)\ge d_i(\mathcal S)$ gives~\eqref{eq:d_lower}.

\subsection{Ramp spectrum}
For the spectrum statement in Section~\ref{subsec:dynamic_sep}, let a $dq$ current component change by $\Delta$ through a linear ramp of duration $\tau$. Its derivative is a rectangular pulse of height $\Delta/\tau$ on $[0,\tau]$, whose Fourier transform has magnitude $\Delta\,|\mathrm{sinc}(\omega\tau/2)|$ with $\mathrm{sinc}(x)=\sin(x)/x$. Dividing by $|\omega|$ gives the ramp's spectrum magnitude away from zero frequency,
\[
\frac{\Delta\,|\mathrm{sinc}(\omega\tau/2)|}{|\omega|}=\frac{2\Delta\,|\sin(\omega\tau/2)|}{\tau\omega^2},
\]
which is approximately $\Delta/|\omega|$ for $|\omega|\ll2/\tau$ and at most $2\Delta/(\tau\omega^2)$ for all $\omega\neq0$.

\section{Network Relation in a POI-Voltage Model}\label{sup:inv}

This section derives~\eqref{eq:q_inversion}. In the quasi-static limit the $q$ component of~\eqref{eq:grid} is
\begin{equation}\label{eqS:q_comp}
v_q=v_{g,q}-R_gi_q-X_gi_d.
\end{equation}
With $v_{g,q}=\bar v_{g,q}$ constant, $i_q=0$, and $X_g\neq0$, solving for $i_d$ gives $i_d=(\bar v_{g,q}-v_q)/X_g$, which is~\eqref{eq:q_inversion}. The relation involves only the network and the source, so it holds whatever the connected set. Any affine regression class that contains $v_q$ as an input admits this relation exactly, with sensitivity $\partial\hat i_d/\partial v_q=-1/X_g$. If a prediction model reproduces the measured current through this relation, then $\hat{\m i}^{(N)}_{\mathcal W}=\m i_{\mathcal W}$ and the least-squares participation ratio $B/A$ in Section~\ref{sup:interval} equals one.

The effect of impedance error on the source reconstruction~\eqref{eq:source_reconstruction} follows by substituting~\eqref{eq:grid} for $\m v$,
\begin{equation}\label{eqS:recon_err}
\hat{\m v}_g=\m v_g+(\hat R_g-R_g)\m i_T+(\hat L_g-L_g)\dot{\m i}_T+\omega_0(\hat L_g-L_g)\m J\m i_T,
\end{equation}
so the residual dependence of $\hat{\m v}_g$ on the facility current is proportional to the impedance error and vanishes when the impedance is exact.

The prediction of Section~\ref{subsec:network_ref} is evaluated in a frame aligned with the reconstructed source voltage. Let $\bar{\m v}_g$ be $\hat{\m v}_g$ filtered with a 10~$\ms$ exponential filter, $\phi_t=\operatorname{atan2}(\bar v_{g,q},\bar v_{g,d})$, and $\m R(\phi)=\cos\phi\,\m I+\sin\phi\,\m J$ the rotation by $\phi$. The constant power term and the correction are computed from $\m R(-\phi_t)\hat{\m v}_g$ and the workload and fitted to $\m R(-\phi_t)\m i_T$ on the calibration records, and the prediction is rotated back by $\m R(\phi_t)$. A rotation of the source voltage, such as a phase jump, then changes $\phi_t$ rather than the inputs once the filter has settled, and when the voltage inputs are restricted to $\|\hat{\m v}_g\|$, relation~\eqref{eq:q_inversion}, which acts through $v_q$, is not available to the fit.

\section{Participation Interval}\label{sup:interval}

This section derives~\eqref{eq:interval}. With $\m i_{\mathcal W},\hat{\m i}^{(N)}_{\mathcal W}\in\RR^{2n}$ the stacked measured and predicted currents, the admissibility condition $\|\m i_{\mathcal W}-a\hat{\m i}^{(N)}_{\mathcal W}\|_2^2\le2n\varepsilon^2$ expands to
\begin{equation}\label{eqS:quad}
Aa^2-2Ba+C\le0,
\end{equation}
with $A=\hat{\m i}^{(N)\top}_{\mathcal W}\hat{\m i}^{(N)}_{\mathcal W}$, $B=\hat{\m i}^{(N)\top}_{\mathcal W}\m i_{\mathcal W}$, and $C=\m i_{\mathcal W}^\top\m i_{\mathcal W}-2n\varepsilon^2$.
For $A>0$ the left side is a convex quadratic in $a$, nonpositive exactly between its roots $(B\pm\sqrt D)/A$ when $D=B^2-AC\ge0$ and nowhere when $D<0$. Intersecting with $[0,1]$ gives the first case of~\eqref{eq:interval}. If $A=0$, then $\hat{\m i}^{(N)}_{\mathcal W}=0$, the condition no longer depends on $a$, and every $a\in[0,1]$ is admissible exactly when $\|\m i_{\mathcal W}\|_2\le\varepsilon\sqrt{2n}$, which gives the second case. All remaining cases are empty.

For the half-width statement, let $\rho^2=\|\m i_{\mathcal W}-(B/A)\hat{\m i}^{(N)}_{\mathcal W}\|_2^2=\m i_{\mathcal W}^\top\m i_{\mathcal W}-B^2/A$ be the residual of the unconstrained least-squares fit. Then
\begin{equation}\label{eqS:D}
D=B^2-A\big(\m i_{\mathcal W}^\top\m i_{\mathcal W}-2n\varepsilon^2\big)=A\big(2n\varepsilon^2-\rho^2\big),
\end{equation}
so the half-width of the unconstrained interval is $\sqrt D/A=\sqrt{2n\varepsilon^2-\rho^2}/\sqrt A$, which approaches $\varepsilon\sqrt{2n}/\|\hat{\m i}^{(N)}_{\mathcal W}\|_2$ when $\rho^2\ll2n\varepsilon^2$.

\section{Feasibility Limit}\label{sup:nose}

This section derives~\eqref{eq:nose}. At unity power factor at the POI the facility current is $\m i_T=(P_T/V^2)\m v$ with $V=\|\m v\|$. Substituting into the quasi-static grid equation and taking the squared norm gives, as in~\eqref{eqS:vg_norm},
\begin{equation}\label{eqS:nose_norm}
\|\m v_g\|^2=V^2+2R_gP_T+\frac{|Z_g|^2P_T^2}{V^2}.
\end{equation}
For $|Z_g|>0$ the right side is a convex function of $V^2>0$ with a unique minimum at $V^2=|Z_g|P_T$, where it equals $2P_T(|Z_g|+R_g)$. A quasi-static operating point with power $P_T$ therefore exists if and only if $\|\m v_g\|^2\ge2P_T(|Z_g|+R_g)$, which is~\eqref{eq:nose}. For $\|\m v_g\|$ above the limit there are two solutions for $V^2$, and the high-voltage branch is the one with $V^2>|Z_g|P_T$.

If instead the converter holds its current in phase with the source voltage, take a frame with $\m v_g=(v_{g,d},0)$ and $\m i_T=(i_d,0)$. The quasi-static grid equation gives $v_d=v_{g,d}-R_gi_d$ and $v_q=-X_gi_d$, so the power is $P_T=\m v^\top\m i_T=v_di_d$ and $v_{g,d}=v_d+R_gP_T/v_d$. For $R_g>0$ the right side has minimum $2\sqrt{R_gP_T}$ over $v_d>0$, which gives the condition $v_{g,d}\ge2\sqrt{R_gP_T}$.

\begin{figure}[t]
\centering
\includegraphics[width=0.9\columnwidth]{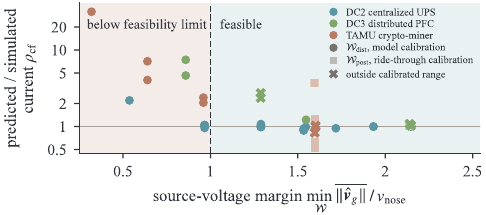}
\caption{Ratio of predicted to simulated all-connected current magnitude across the feasibility limit~\eqref{eq:nose}, first 37 events. Crosses mark windows outside the calibrated range.}
\label{fig:cs_feasibility}
\end{figure}

\section{Facilities With Multiple Protection Zones}\label{subsec:zones}

When a facility has several protection zones with different dynamics, a single participation value is only an approximation. A closed-loop facility model can instead be evaluated for each candidate connected set so that the POI voltage and the response of the remaining zones are recomputed consistently. This requires the workload of each candidate set. If we only have aggregate workload data, it must be allocated from the total, which still includes the requested workload of any disconnected zone. Let $P_k=\bar P_k+\tilde P_k$, where the pre-disturbance mean $\bar P_k$ is proportional to $P_k^0$ and $\tilde P_k$ is a zero-mean fluctuation, and let $0<c(\mathcal S)<1$. Allocating the aggregate in proportion to nominal power gives the error
\begin{equation}\label{eq:telemetry_gap}
c(\mathcal S)\,w_T-w_{\mathcal S}
=c(\mathcal S)\sum_{k\notin\mathcal S}\tilde P_k-\big(1-c(\mathcal S)\big)\sum_{k\in\mathcal S}\tilde P_k,
\end{equation}
which is zero exactly when the connected and disconnected zones have the same total fluctuation per unit nominal power (Section~\ref{sup:zone}).

With zone-level workload data, $w_{\mathcal S}$ is observed directly and this error disappears. Let $\hat{\m i}_{\mathcal W}(\mathcal S')$ be the predicted current when only the zones in $\mathcal S'$ are connected. The connected sets consistent with the data are then
\begin{equation}\label{eq:subset_family}
\mathfrak S(\varepsilon)=\big\{\mathcal S'\subseteq\{1,\ldots,N\}\ \big|\ \|\m i_{\mathcal W}-\hat{\m i}_{\mathcal W}(\mathcal S')\|_2\le\varepsilon\sqrt{2n}\big\}.
\end{equation}
Enumeration requires up to $2^N$ simulations which is only practical for a small number of zones. Large facilities with many zones would likely require a more structured search instead. A single admissible set identifies which zones disconnected. If several zones are electrically indistinguishable and have proportional workloads, more than one set remains admissible and the data cannot tell those sets apart. Zone identification with~\eqref{eq:subset_family} was not evaluated in this work.

\section{Allocation Error With Aggregate Workload}\label{sup:zone}

This section derives~\eqref{eq:telemetry_gap}. Let $P_k=\bar P_k+\tilde P_k$ with $\bar P_k=\gamma P_k^0$ for a common $\gamma>0$ and zero-mean fluctuations $\tilde P_k$. The zones in $\mathcal S$ draw $w_{\mathcal S}=\sum_{k\in\mathcal S}P_k$, while a prediction driven by the aggregate workload data and scaled to the connected nominal power fraction is driven by $c(\mathcal S)\,w_T$ with $w_T=\sum_kP_k$. By the definition of $c(\mathcal S)$ in~\eqref{eq:capacity_count}, $c(\mathcal S)\sum_k\bar P_k=\gamma c(\mathcal S)\sum_kP_k^0=\gamma\sum_{k\in\mathcal S}P_k^0=\sum_{k\in\mathcal S}\bar P_k$, so the means cancel in the difference and
\begin{align}\label{eqS:alloc}
c(\mathcal S)\,w_T-w_{\mathcal S}
&=c(\mathcal S)\sum_k\tilde P_k-\sum_{k\in\mathcal S}\tilde P_k\nonumber\\
&=c(\mathcal S)\sum_{k\notin\mathcal S}\tilde P_k-\big(1-c(\mathcal S)\big)\sum_{k\in\mathcal S}\tilde P_k,
\end{align}
which is~\eqref{eq:telemetry_gap}. For $0<c(\mathcal S)<1$, the right side is zero if and only if $\sum_{k\notin\mathcal S}\tilde P_k/(1-c(\mathcal S))=\sum_{k\in\mathcal S}\tilde P_k/c(\mathcal S)$. Since $(1-c(\mathcal S))\sum_kP_k^0=\sum_{k\notin\mathcal S}P_k^0$ and $c(\mathcal S)\sum_kP_k^0=\sum_{k\in\mathcal S}P_k^0$, this is the equality of fluctuations per unit nominal power between the disconnected and connected zones. For independent fluctuations with variances $\sigma_k^2$, the variance of~\eqref{eqS:alloc} is $c(\mathcal S)^2\sum_{k\notin\mathcal S}\sigma_k^2+(1-c(\mathcal S))^2\sum_{k\in\mathcal S}\sigma_k^2$. The identity is an input-allocation identity and does not assume a common workload-to-current operator across zones. With zone-level data, $w_{\mathcal S}$ is observed directly and the error is absent.
\end{document}

%% file: preamble.tex
\usepackage{balance}
\usepackage{graphicx}
\usepackage{amsmath,amsfonts,amssymb,amsthm}
\usepackage{bm}
\usepackage{array}
\usepackage{booktabs}
\usepackage{multirow}
\usepackage{cite}
\usepackage{url}
\usepackage{xcolor}
\usepackage{enumitem}
\usepackage{mathtools}
\usepackage{stfloats}
\usepackage{tikz}
\usetikzlibrary{arrows.meta,positioning,calc,shapes.geometric}
\usepackage{lettrine}
\usepackage[font=footnotesize]{caption}
\usepackage[colorlinks=true,linkcolor=blue,urlcolor=black,citecolor=blue,anchorcolor=blue]{hyperref}
\newcommand{\m}{\boldsymbol}
\newcommand{\RR}{\mathbb{R}}
\newcommand{\dd}{\mathrm{d}}
\newcommand{\pu}{\mathrm{p.u.}}
\newcommand{\Hz}{\mathrm{Hz}}
\newcommand{\ms}{\mathrm{ms}}
\newcommand{\col}{\operatorname{col}}

\DeclareMathOperator{\dist}{dist}

\allowdisplaybreaks[4]

%% file: tables/Table_CS1_testbeds.tex
\begin{table*}[t]
\centering
\caption{Test Systems Used in the Case Studies}
\label{tab:cs_testbeds}
\footnotesize
\setlength{\tabcolsep}{5pt}
\renewcommand{\arraystretch}{1.10}
\begin{tabular}{@{}>{\raggedright\arraybackslash}p{2.8cm}>{\raggedright\arraybackslash}p{5.0cm}>{\raggedright\arraybackslash}p{3.6cm}>{\raggedright\arraybackslash}p{5.0cm}@{}}
\toprule
Test system & Facility representation & Grid boundary & Electrical loss represented \\
\midrule
PNNL DML DC2~\cite{ross2026dml} &
700~MW at 230~kV. Centralized double-conversion UPS units with battery-backed DC links supplying the computational load. &
$R_g=0.0131$, $X_g=0.131$~$\pu$ on 1000~MVA, $X/R=10$. &
The UPS input stage can stop drawing grid current during low voltage, and a separate support branch can open. \\
\addlinespace[2pt]
PNNL DML DC3~\cite{ross2026dml} &
700~MW at 230~kV. Distributed single-phase power-factor-correction (PFC) front ends with local DC links and battery converters. &
$R_g=0.0131$, $X_g=0.131$~$\pu$ on 1000~MVA, $X/R=10$. &
Selected groups of front-end converters can stop drawing grid current during low voltage, and a support branch can open. \\
\addlinespace[2pt]
TAMU/ERCOT crypto-miner~\cite{wang2025crypto} &
300~MW at 138~kV. Average-value boost-PFC front ends supplying constant-power DC loads without battery storage. &
$R_g=0.020$, $X_g=0.200$~$\pu$ on 300~MVA, $X/R=10$. &
Voltage ride-through protection trips the affected miner units and opens their breakers. \\
\addlinespace[2pt]
Versions with separate zone settings &
The three test systems with zones given different protection settings: DC2 UPS (91\%) and support (9\%) branches; DC3 one PFC group per phase; TAMU three equal groups of miner units. &
As above. &
Zones trip separately. Undervoltage thresholds between 0.70 and 0.80~$\pu$ with delays between 50 and 120~ms let a sag disconnect some zones and not others. \\
\bottomrule
\end{tabular}
\vspace{-0.4cm}
\end{table*}

%% file: tables/Table_CS2_results.tex
\begin{table*}[t]
\centering
\caption{Results for the PSCAD Evaluation Events. Errors, classification errors, and set counts are over admitted events and include events with $\mathcal A(\varepsilon)=\emptyset$. An estimate $\hat a<0.95$ classifies an event as a loss. Load errors are mean absolute errors in percent of the pre-event power.}
\label{tab:cs_results}
\scriptsize
\setlength{\tabcolsep}{3pt}
\begin{tabular}{@{}llcccccccccccc@{}}
\toprule
 & & & & & \multicolumn{2}{c}{$|\hat a-a^\star|$} & \multicolumn{2}{c}{Classification errors} & \multicolumn{3}{c}{Consistency set $\mathcal A(\varepsilon)$} & \multicolumn{2}{c}{Load error (\%)} \\
\cmidrule(lr){6-7}\cmidrule(lr){8-9}\cmidrule(lr){10-12}\cmidrule(lr){13-14}
Window, calibration & Test system & Events & Admitted & $\varepsilon$ ($\pu$) & Mean & Max & False loss & Missed loss & Contains & Empty & Excludes & Disconnection & Workload \\
\midrule
\multicolumn{14}{@{}l}{\emph{First 37 events}} \\
$\mathcal W_{\rm dist}$, facility model & DC2 centralized UPS & 23 & 16 & 0.116 & 0.012 & 0.045 & 0/8 & 0/8 & 8 & 8 & 0 & 3.1 & 2.8 \\
 & DC3 distributed PFC & 7 & 1 & 0.517 & 0.104 & 0.104 & 0/0 & 0/1 & 1 & 0 & 0 & 22.8 & 20.8 \\
 & TAMU crypto-miner & 7 & 0 & 0.061 & -- & -- & 0/0 & 0/0 & 0 & 0 & 0 & -- & -- \\
\addlinespace[2pt]
$\mathcal W_{\rm post}$, ride-through records & DC2 centralized UPS & 23 & 22 & 0.018 & 0.014 & 0.034 & 0/7 & 0/15 & 7 & 15 & 0 & 2.2 & 5.5 \\
 & DC3 distributed PFC & 7 & 4 & 0.018 & 0.014 & 0.018 & 0/0 & 0/4 & 0 & 4 & 0 & 1.9 & 8.0 \\
 & TAMU crypto-miner & 7 & 4 & 0.084 & 0.000 & 0.000 & 0/0 & 0/4 & 4 & 0 & 0 & 97.2 & 97.1 \\
\addlinespace[2pt]
\addlinespace[2pt]
\multicolumn{14}{@{}l}{\emph{60 partial-trip events}} \\
$\mathcal W_{\rm dist}$, facility model & DC2, two zones & 20 & 20 & 0.088 & 0.039 & 0.237 & 0/7 & 0/13 & 6 & 14 & 0 & 5.2 & 6.2 \\
 & DC3, per-phase zones & 20 & 14 & 0.275 & 0.063 & 0.163 & 0/4 & 0/10 & 10 & 4 & 0 & 6.9 & 8.9 \\
 & TAMU, three groups & 20 & 12 & 0.063 & 0.018 & 0.074 & 0/7 & 0/5 & 6 & 6 & 0 & 3.9 & 2.9 \\
\addlinespace[2pt]
$\mathcal W_{\rm post}$, ride-through records & DC2, two zones & 20 & 20 & 0.019 & 0.023 & 0.044 & 0/3 & 0/17 & 1 & 18 & 1 & 3.4 & 7.4 \\
 & DC3, per-phase zones & 20 & 14 & 0.020 & 0.013 & 0.019 & 0/4 & 0/10 & 4 & 10 & 0 & 1.4 & 4.9 \\
 & TAMU, three groups & 20 & 12 & 0.090 & 0.042 & 0.102 & 1/7 & 0/5 & 9 & 0 & 3 & 6.2 & 5.2 \\
\bottomrule
\end{tabular}
\vspace{-0.5cm}
\end{table*}